\documentclass[article]{JHEP3} 

\usepackage[section] {placeins}
\usepackage{epsfig,multicol}
\usepackage{amsmath}
\usepackage{graphicx}
\usepackage[latin1]{inputenc}
\usepackage{amssymb}
\usepackage{amsmath}
\usepackage{verbatim}
\usepackage{epsfig}
\usepackage{multirow}
\usepackage{booktabs}
\newcommand\fverb{\setbox\pippobox=\hbox\bgroup\verb}
\newcommand\fverbdo{\egroup\medskip\noindent%
                      \fbox{\unhbox\pippobox}\ }

\newcommand\fverbit{\egroup\item[\fbox{\unhbox\pippobox}]}
\newbox\pippobox
\catcode`~=11 
\newcommand{\urltilde}{\kern -.15em\lower .7ex\hbox{~}\kern .04em}
\catcode`~=13 

\newcommand{\eg}{{\it e.g.}}

\newcommand{\pythia}{{\sc Pythia}}

\newcommand{\herwigpp}{{\sc Herwig++}}
\newcommand{\herwig}{{\sc Herwig}}

\newcommand{\madloop}{{\sc MadLoop}}
\newcommand{\cuttools}{{\sc CutTools}}
\newcommand{\madfks}{{\sc MadFKS}}
\newcommand{\mcatnlo}{{\sc MC@NLO}}
\newcommand{\amcatnlo}{a{\sc MC@NLO}}
\newcommand{\amcatlo}{a{\sc MC@LO}}
\newcommand{\Wbb}{$\ell\nu b\bar b$}
\newcommand{\Zbb}{$\ell^+ \ell^- b\bar b$}
\newcommand{\Wbbs}{$Wb\bar{b}$}
\newcommand{\Zbbs}{$Zb\bar{b}$}

\newcommand{\be}{\begin{equation}}
\newcommand{\ee}{\end{equation}}
\newcommand{\ba}{\begin{eqnarray}}
\newcommand{\ea}{\end{eqnarray}}
\newcommand{\bt}{\begin{tabular}}
\newcommand{\et}{\end{tabular}}
\newcommand{\bfig}{\begin{figure}}
\newcommand{\efig}{\end{figure}}

\newcommand\sss{\scriptscriptstyle}

\newcommand{\pt}{p_{\sss T}}
\newcommand{\Pt}{P_{\sss T}}
\newcommand{\kt}{k_{\sss T}}

\newcommand{\abs}[1]{\left|#1\right|}

\newcommand\muF{\mu_{\sss F}}
\newcommand\muR{\mu_{\sss R}}

\preprint{
 CERN-PH-TH/2011-147 \\
 CP3-11-20 \\
 NSF-KITP-11-114 \\
 ZH-TH 13/11
 }
\title{$W$ and $Z/\gamma^*$ boson production in association with a
bottom-antibottom pair}

\author{Rikkert Frederix\\
Institut f\"ur Theoretische Physik, Universit\"at Z\"urich,
Winterthurerstrasse 190,\\ CH-8057 Z\"urich, Switzerland\\
KITP, University of California Santa Barbara, CA 93106-4030, USA 
}
\author{Stefano Frixione%
  \thanks{On leave of absence from INFN, Sezione di Genova, Italy.}\\
  PH Department, TH Unit, CERN, CH-1211 Geneva 23, Switzerland\\
  ITPP, EPFL, CH-1015 Lausanne, Switzerland
}
\author{Valentin Hirschi\\
  ITPP, EPFL, CH-1015 Lausanne, Switzerland
}
\author{Fabio Maltoni\\
  Centre for Cosmology, Particle Physics and Phenomenology (CP3)\\
  Universit\'{e} catholique de Louvain,  B-1348 Louvain-la-Neuve, Belgium
}
\author{Roberto Pittau\\
  Departamento de F\'\i sica Te\'orica y del Cosmos y CAFPE, 
  Universidad de Granada\\
  PH Department, TH Unit, CERN, CH-1211 Geneva 23, Switzerland\\
  KITP, University of California Santa Barbara, CA 93106-4030, USA 
}
\author{Paolo Torrielli\\
  ITPP, EPFL, CH-1015 Lausanne, Switzerland
}

\abstract{We present a study of \Wbb\ and
\Zbb\ production at hadron colliders.
Our results, accurate to the next-to-leading order in QCD, are based on 
automatic matrix-element calculations performed by \madloop\ and \madfks,
and are given at both the parton level, and after the matching with
the \herwig\ event generator, achieved with \amcatnlo. We retain
the complete dependence on the bottom-quark mass, and include exactly 
all spin correlations of final-state leptons. We discuss the cases of
several observables at the LHC which highlight the importance of 
accurate simulations.}

\keywords{LHC, NLO, Monte Carlo}

\begin{document}


\section{Introduction}
\label{sec:intro}

The discovery and identification of new degrees of freedom and interactions at
high-energy colliders relies on the detailed understanding of Standard Model
(SM) background processes. Prominent among these is the production of
electroweak bosons ($W,Z$) in association with jets, of which one or more 
possibly contain bottom quarks. The prime example is the observation of the
top quark at the Fermilab Tevatron collider, produced either in
pairs~\cite{Abe:1995hr,Abachi:1995iq} or
singly~\cite{Abazov:2009ii,Aaltonen:2009jj}, since both of 
these mechanisms typically lead to $W$ plus $b$-jets signatures.  
Other crucial examples involve the search for a SM
Higgs in association with vector bosons ($WH/ZH$),
with the subsequent Higgs decay into a $b\bar b$ pair, a sought-for discovery
channel both at the Tevatron~\cite{:2010ar} and at the
LHC~\cite{Ball:2007zza,Aad:2009wy}. Finally, in models which feature an
extended Higgs sector, such as the MSSM or more generally a two-Higgs doublet
model, a typical Higgs discovery channel is through $Hb\bar b$ and $Ab\bar b$ 
final states, with an $H/A \to\tau^+\tau^-$ decay. In this case, the SM 
process \Zbb\ can provide an important reference measurement.

Next-to-leading-order (NLO) QCD calculations for the production of a vector
boson in association with jets have by now quite a successful record. Accurate
predictions for $W$ plus up to four light jets~\cite{Arnold:1988dp,
Campbell:2002tg,Berger:2009ep,Berger:2009zg,Melnikov:2009wh,Berger:2010zx}
and for $Z$ plus up to three light
jets~\cite{Ellis:1981hk,Arnold:1988dp,Campbell:2002tg,Berger:2010vm} have
become available in the past few years. Associated production with heavy 
quarks, and in particular with bottom quarks, has been studied using various
approximations. The \Wbbs\ and \Zbbs\ processes have been 
calculated at the NLO for the
first time in refs.~\cite{Ellis:1998fv} and~\cite{Campbell:2000bg}
respectively, by setting the bottom-quark mass equal to zero.
Such calculations can be used only for observables that contain at least 
two $b$-jets. The same processes have been considered again
in refs.~\cite{Cordero:2006sj,FebresCordero:2008ci,Cordero:2009kv}, 
where a non-zero bottom-quark mass has been used; however, the matrix
elements still involved on-shell vector bosons, thereby neglecting
spin correlations of the leptons emerging from $W$ and $Z$ decays.
In the case of $W$ production, this limitation has been recently lifted 
in ref.~\cite{Badger:2010mg}, which presents the NLO calculation for the
leptonic process \Wbb.
Other NLO calculations for final states with one $b$-jet, $Wb$ and 
$Zb$~\cite{Campbell:2008hh,Campbell:2003dd},
and one $b$-jet plus a light jet, $Wbj$ and
$Zbj$~\cite{Campbell:2006cu,Campbell:2005zv}, are also available in the
five-flavour scheme. All such calculations have played a role 
and/or have been extensively compared to the data collected at the
Tevatron~\cite{Acosta:2001ct,Aaltonen:2008mt,Aaltonen:2009qi,Abazov:2010ix},
and now start to be considered in LHC analyses as well~\cite{wzcms,wzatlas}.

In this paper we present a calculation of \Wbb\ production
that includes NLO QCD corrections (analogous to that of
ref.~\cite{Badger:2010mg}), and the first calculation at the NLO of 
\Zbb\ production with massive bottom quarks; we retain the full
spin correlations of the final-state leptons\footnote{In the rest of
this paper, \Wbb\ and \Zbb\ production as predicted by our simulations
may also be denoted by \Wbbs\ and \Zbbs\ respectively.}. 
Furthermore, we match both of these results to the \herwig\ event generator by
adopting the MC@NLO formalism~\cite{Frixione:2002ik}. Therefore, our results
include all the relevant features which are important in experimental
analyses, and can be used in order to obtain NLO predictions for a large class
of observables, including those with zero, one and two $b$-jets. All aspects 
of the calculations are fully automated and analogous to the calculation 
recently appeared for $Ht\bar t/A t\bar t $ production~\cite{Frederix:2011zi}. 
One-loop amplitudes
are evaluated with \madloop~\cite{Hirschi:2011pa}, whose core is the OPP 
integrand reduction method~\cite{Ossola:2006us} as implemented in
\cuttools~\cite{Ossola:2007ax}.  Real contributions and the corresponding
phase-space subtractions, achieved by means of the FKS 
formalism~\cite{Frixione:1995ms}, as well as their combination with the 
one-loop and Born results and their subsequent integration, are performed by
\madfks~\cite{Frederix:2009yq}. The MC@NLO matching is also fully automated,
and allows us to simulate for the first time \Wbb\ and \Zbb\ production with 
NLO accuracy, including {\em exactly} spin correlations, off-shell 
and interference effects, and hadron-level final states. All the 
computations are integrated in a single software framework, which 
we have dubbed \amcatnlo\ in ref.~\cite{Frederix:2011zi}.
We point out that the on-shell-$W$ result of ref.~\cite{Cordero:2006sj} 
has recently been matched to showers in ref.~\cite{Oleari:2011ey} in the 
framework of the POWHEG box~\cite{Alioli:2010xd}.

The phenomenology of \Wbb\ and \Zbb\ final states is very rich and
in fact transcends the prosaic role of background for Higgs or top-quark
physics. Thanks to the \amcatnlo\ implementation, several QCD issues
interesting on their own can now be addressed theoretically and the results
efficiently compared to experiments.  In this work we limit ourselves to
providing new evidence that reliable and flexible predictions for 
the \Wbb\ and \Zbb\ processes need to include:

\begin{itemize}
\item NLO corrections;
\item bottom quark mass effects;
\item spin-correlation and off-shell effects;
\item showering and hadronisation.
\end{itemize}

\noindent
Detailed studies of these processes as backgrounds to specific signals,
such as single-top and $Hb\bar b$ or $Ab\bar b$ production respectively, 
are left to forthcoming investigations.

This paper is organised as follows. In the next
section we present several distributions relevant to \Wbb\ and \Zbb\
production at the LHC, and report the results for total rates at both
the Tevatron and the LHC. By working with a non-zero bottom mass, we are
able to obtain predictions for the cases in which one or two $b$'s are
not observed, and can thus have arbitrarily small transverse momenta.
We also give one example of the comparisons,
at the level of hadronic final states, between the $HW$ and $HZ$
signals and their respective irreducible backgrounds which we
have computed in this paper. We draw our conclusions in
sect.~\ref{sec:con}.

\section{Results}
\label{sec:inclusive}

\begin{figure}[t]
\centering
\includegraphics[width=0.7\textwidth]{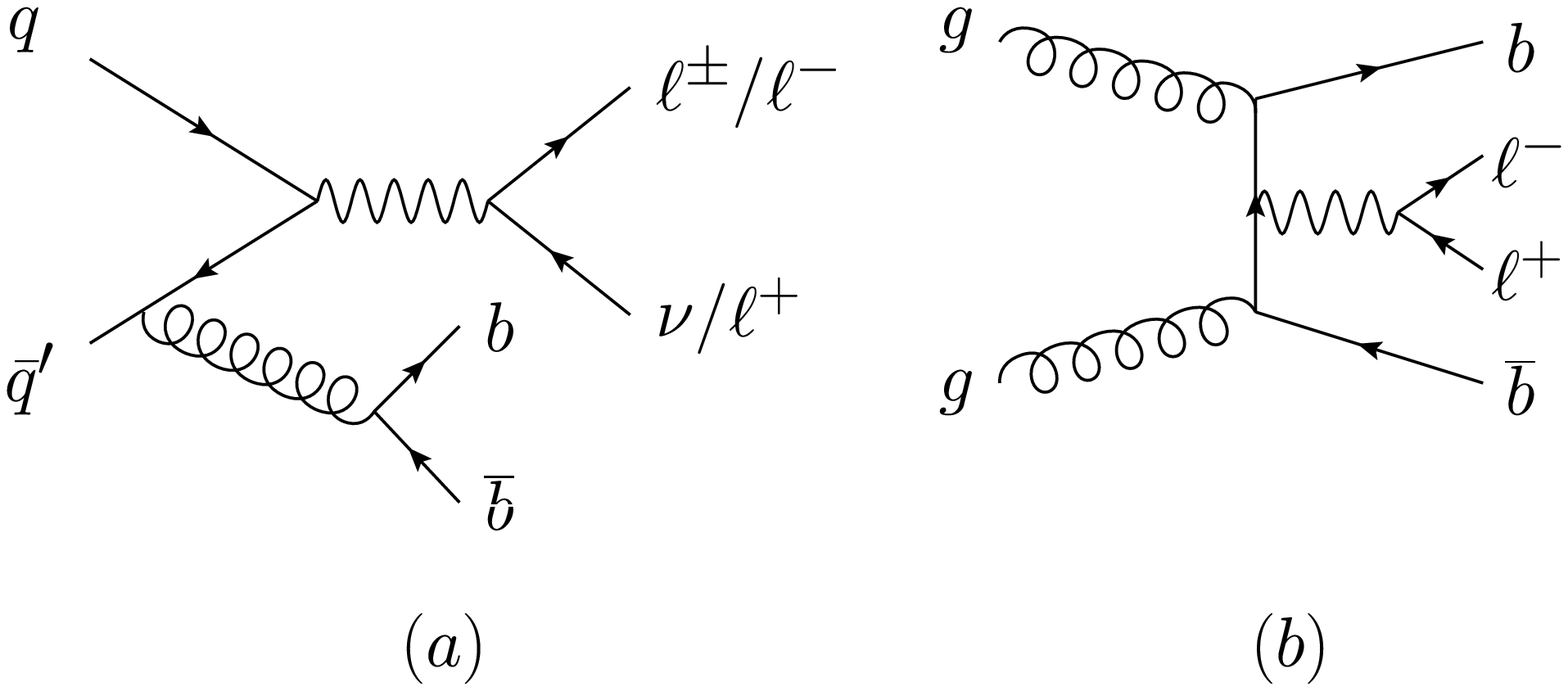}
\caption{Representative diagrams contributing to \Wbb\ and \Zbb\ production at
the leading order. \Wbb\ production can proceed only via a $q\bar{q}^\prime$ 
channel, diagram (a). For \Zbb\ production the $q\bar{q}$ channel, 
diagram (a), is dominant at the Tevatron, while the $gg$ channel, 
diagram (b), largely dominates at the LHC.}
\label{fig:diagrams}
\end{figure}
At the leading order (LO) in QCD \Wbb\ and \Zbb\ production at 
hadron colliders proceed through different channels. Both final states 
can be obtained via a Drell-Yan-type mechanism, i.e., 
$q\bar{q}^{(\prime)}$ annihilation in
association with a gluon splitting in a $b \bar b$ pair, see
fig.~\ref{fig:diagrams}(a). \Zbbs, however, can also be produced by
gluon fusion, see fig.~\ref{fig:diagrams}(b), a channel that at the LO 
contributes a 30\% of the total rate at the Tevatron,
but turns out to be the dominant one (80\%) at the LHC, owing to 
the larger gluon luminosity
there. As we shall see in the following, the fact that \Wbbs\ and 
\Zbbs\ production are dominated by different channels at the LHC 
leads to important differences in the kinematical properties of 
final states, and in particular of $b$-jets.

\begin{table}
\begin{center}
\begin{tabular}{ll|ll}\toprule
Parameter & value & Parameter & value
\\\midrule
$m_{Z}$ & 91.118 & $\alpha^{-1}$ & 132.50698 
\\
$m_{W}$ & 80.419 & $G_F$         & $1.16639\!\cdot\!10^{-5}$
\\
$m_b$        & 4.5    & ${\rm CKM}_{ij}$ & $\delta_{ij}$
\\
$m_t$      & 172.5  & $\Gamma_Z$ & 2.4414 
\\
$\alpha_s^{({\rm LO},4)}(m_{Z})$  & 0.133551 & $\Gamma_W$ & 2.0476 
\\
$\alpha_s^{({\rm NLO},4)}(m_{Z})$ & 0.114904 & 
\\\bottomrule
\end{tabular}
\end{center}
\caption{\label{tab:params}
  Settings of physical parameters used in this work,
  with dimensionful quantities given in GeV.
}
\end{table}

We start by presenting results for the total cross sections at both
the Tevatron, $\sqrt{s}=1.96$ TeV, and the LHC, $\sqrt{s}=7$ TeV;
the \Wbb\ results are the sums of the $\ell^+ \nu b\bar b$
and $\ell^-\bar{\nu} b\bar b$ ones (due to virtual-$W^+$ and $W^-$
production respectively).
In our computations we have set the lepton masses equal to zero,
and is therefore not necessary to specify their flavour, which we
generically denote by $\ell$ (for the charged leptons) and
$\nu$ (for the neutrinos); we always quote results for one flavour. 
For the numerical analysis we have chosen:
\ba
\muF^2 = \muR^2 &=& 
m_{\ell\ell^\prime}^2+\pt^2{(\ell\ell^\prime)} + 
\frac{m_b^2+\pt^2(b)}{2}+
\frac{m_b^2+\pt^2({\bar b})}{2}\,,
\ea
with $\ell\ell^\prime=\ell\nu$ and $\ell\ell^\prime=\ell^+\ell^-$
in the case of \Wbbs\ and \Zbbs\ production respectively; the 
value of the $b$-quark mass is that of the pole mass, \mbox{$m_b=4.5$~GeV}.
We have used LO and NLO MSTW2008 four-flavour parton 
distribution functions~\cite{Martin:2009iq} for the corresponding cross
sections, and the SM-parameter settings can be found in
table~\ref{tab:params}. Given that the primary aim of this paper
is not that of presenting precise comparisons with data, but rather
that of studying the defining features of the production mechanisms, 
in the CKM matrix (relevant to the \Wbbs\ results) we have neglected
off-diagonal terms: this cannot change the conclusions we shall
arrive at, but helps reduce the computing time. It should be clear
that this is not a limitation of the code, since a non-diagonal CKM
matrix can simply be given in input if one so wishes.
Our runs are fully inclusive and no cuts are applied
at the generation level, except for $m_{\ell^+ \ell^-}>30$ GeV in the 
\Zbb\ sample. The predicted production rates
at the Tevatron and at the LHC are given in
table~\ref{tab:xsec} where, for ease of reading, we also show the fully
inclusive $K$ factors. The contribution of the $gg\to Zb\bar{b}+X$ channels 
is clearly visible in these results: at the Tevatron 
\mbox{$\sigma(\ell^+ \ell^- b\bar b)/\sigma(\ell\nu b\bar b)$}
is quite small (and of the same order of the ratio of
the fully-inclusive cross sections $\sigma(Z)/\sigma(W)$), whereas at the LHC
\Zbb\ and \Wbb\ differ only by a factor of two.

\renewcommand{\arraystretch}{1.7}
\begin{table}[t]
\begin{center}
\begin{tabular}{|c|ccc|ccc|}
\hline
 & \multicolumn{6}{c|}{Cross section (pb)}\\

\cline{2-7}  
    & \multicolumn{3}{c|}{Tevatron $\sqrt{s}=$1.96 TeV} 
    & \multicolumn{3}{c|}{LHC $\sqrt{s}=$7 TeV} \\

 & LO &NLO &$K$ factor& LO & NLO & $K$ factor \\
\hline
$\phantom{aa}$\Wbb$\phantom{aa}$ & 4.63  & 8.04  & 1.74  
                                 & 19.4  & 38.9  & 2.01  \\
\Zbb\                            & 0.860 & 1.509 & 1.75 
                                 &  9.66 & 16.1  & 1.67  \\
\hline
\end{tabular}
\end{center}
\caption{Total cross sections for \Wbb\ and \Zbb\ production
at the Tevatron ($p\bar p$ collisions at
$\sqrt{s}= 1.96$ TeV) and the LHC ($pp$ collisions at $\sqrt{s}= 7$ TeV), to
LO and NLO accuracy. These rates are relevant to one lepton flavour, and the
results for \Wbb\ production are the sums of those for $\ell^+ \nu b\bar b$
and $\ell^-\bar{\nu} b\bar b$ production. The integration uncertainty is 
always well below 1\%.}
\label{tab:xsec}
\end{table}

We now study the impact of NLO QCD corrections on differential 
distributions, at both the parton level and after showering 
and hadronisation, and in doing so we limit ourselves to the
case of the LHC, where the kinematical differences between \Wbbs\ 
and \Zbbs\ production are more evident.
The parton shower in \amcatnlo\ has been performed with fortran
\herwig~\cite{Marchesini:1991ch,Corcella:2000bw,Corcella:2002jc}, version
6.520\footnote{Automation of the matching to parton shower in the \mcatnlo\
formalism to \herwigpp~\cite{Bahr:2008pv} and to
\pythia~\cite{Sjostrand:2006za} (see refs.~\cite{Frixione:2010ra} 
and~\cite{Torrielli:2010aw} respectively) is currently under way.}.
\begin{figure}[t]
\centering
\includegraphics[width=0.6\textwidth]{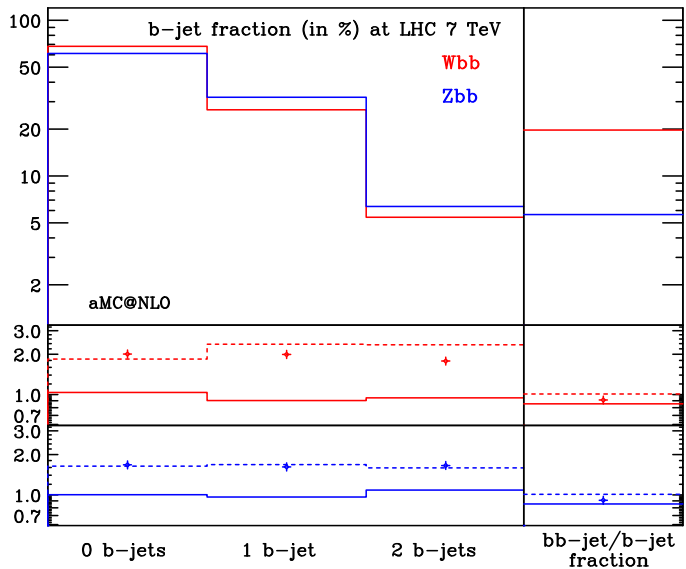}
\caption{Fractions of events (in percent) that contain: zero $b$-jets,
  exactly one $b$-jet, and exactly two $b$-jets. The rightmost bin displays
  the fraction of $b$-jets which are $bb$-jets. The two insets show 
  the ratio of the \amcatnlo\ results over the corresponding NLO (solid),
  \amcatlo\ (dashed), and LO (symbols) ones, separately for \Wbbs\ 
  (upper inset) and \Zbbs\ (lower inset) production.}
\label{fig:rates}
\end{figure}

We start by summarizing our results for $b$-jet rates. Jets are reconstructed 
at the particle level. In the case of MC simulations, this means giving all 
final-state stable hadrons\footnote{In order to simplify the \herwig\ analyses,
weakly-decaying $B$ hadrons are set stable.} in input to the jet algorithm. 
We adopt the anti-$\kt$ jet clustering algorithm~\cite{Cacciari:2008gp}
with $R=0.5$, and require each jet to have $\pt(j)>20$~GeV and 
$|\eta(j)|<2.5$. A $b$-jet is then defined as a jet that contains 
{\em at least} one $b$-hadron; a $bb$-jet is a jet that contains at least 
two $b$-hadrons (hence, a $bb$-jet is also a $b$-jet).
This implies that we make no distinction between the $b$ quark and antiquark
contents of a jet. We point out that at least another definition of $b$-jets
exists~\cite{Banfi:2006hf} which has a better behaviour in the
$m_b\to 0$ limit, in the sense that it gives (IR-safe) results consistent 
with the naive picture of ``quark'' and ``gluon'' jets. In practice, this 
is relevant only in the $\pt\gg m_b$ limit. Since this region is not our 
primary interest in this paper, we stick to the usual definition; however,
it should be obvious that any jet definition can be used in our framework.

In fig.~\ref{fig:rates} we present $b$-jet rates, as
the fractions of events that contain zero, exactly one,
or exactly two $b$-jet(s). In the case of MC-based simulations, there are
also events with more than two $b$-jets and more than one $bb$-jet,
but they give a relative contribution to the total rate equal to about
0.4\% (for \Wbbs) and 0.6\% (for \Zbbs), and are therefore
not reported here. The rightmost bin of fig.~\ref{fig:rates} shows the
fraction of $b$-jets which are $bb$-jets. There is an inset for each 
of the two histograms shown in the upper part of fig.~\ref{fig:rates}. 
Each of the insets presents three curves, obtained by computing the ratio 
of the \amcatnlo\ results over the NLO (solid), \amcatlo\footnote{We call 
\amcatlo\ the analogue of \amcatnlo, in which the short-distance cross 
sections are computed at the LO rather than at the NLO. Its results 
are therefore equivalent to those one would obtain by using, \eg, 
{\sc MadGraph/MadEvent}~\cite{Alwall:2007st} interfaced to showers.}
(dashed), and LO (symbols) corresponding ones.
The $b$-jet fractions are fairly similar for \Wbbs\ and \Zbbs\ production,
and the effects of the NLO corrections are consistent with the 
fully-inclusive $K$ factors. On the other hand,
the $bb$-jet contribution to the $b$-jet rate is seen to be more than
three times larger for \Wbb\ than for \Zbb\ final states.
This fact is again related to the different mechanisms for the 
production of a $b\bar{b}$ pair in the two processes considered here.
At variance with the case of \Wbb\ production, in a \Zbb\ final state
the two $b$'s may come from the separate branchings of two initial-state 
gluons, and thus the probability of them ending in the same jet is much
smaller than in the case of a $g\to b\bar{b}$ final-state branching.
We conclude this discussion by pointing out that the zero- and
one-$b$-jet rates can only be obtained with a non-zero $b$-quark mass,
since the one or two ``untagged'' $b$'s must be integrated down to $\pt=0$, 
and hence $m_b\ne 0$ is required in order to screen initial-state collinear
divergences. This fact is a severe test condition for the computer
programs used in the computations, because it may induce numerical
instabilities. We stress that we did not impose low-$\pt$ cuts on any
of the final-state particles in \madloop, \madfks, and in the generation
of hard events in \amcatnlo; our results are therefore completely unbiased,
which is what gives us the possibility of computing quantities such as 
those reported in table~\ref{tab:xsec} and in fig.~\ref{fig:rates}. 

We now turn to studying differential distributions, and start
by considering those defined in terms of final-state leptons.
Observables sensitive to the hadronic activity of the events,
be either relevant to $b$-jets or to $B$-hadrons, will follow later.
In the case of MC-based simulations, several leptons can appear
in the final state. We use the MC-truth information to select the
two which emerge from the hard process, and we shall simply refer to 
them as ``the leptons'' henceforth. A more realistic analysis may 
select leptons on the basis of their hardness, measured e.g. with
their $\pt$'s; in practice, for the processes we are considering
here (and thanks to the fact that the $b$-hadrons have been set stable)
the two approaches are equivalent.

\begin{figure}[t]
\centering
\includegraphics[width=0.49\textwidth]{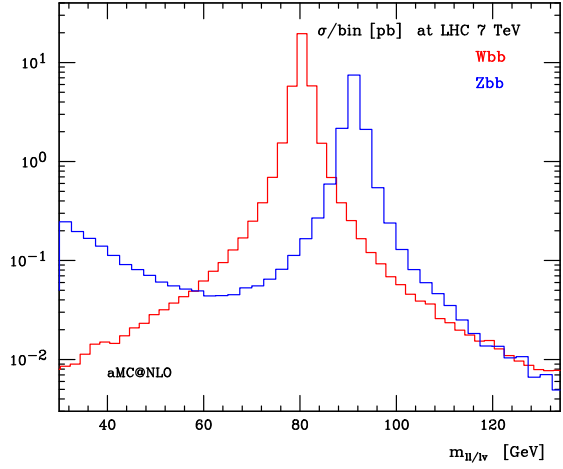}
\caption{Invariant mass distribution of final-state lepton pairs,
as predicted by \amcatnlo.}
\label{fig:M_ee_2}
\end{figure}
In fig.~\ref{fig:M_ee_2} the invariant mass of the final-state 
lepton pairs is shown. The effect of the $\gamma^*$ contribution 
to \Zbb\ production is clearly visible at small invariant 
masses. In this plot, with limit ourselves to presenting only
the \amcatnlo\ results, since the other simulations give results
which are essentially identical to the present ones.

\begin{figure}[t]
\centering
\includegraphics[width=0.49\textwidth]{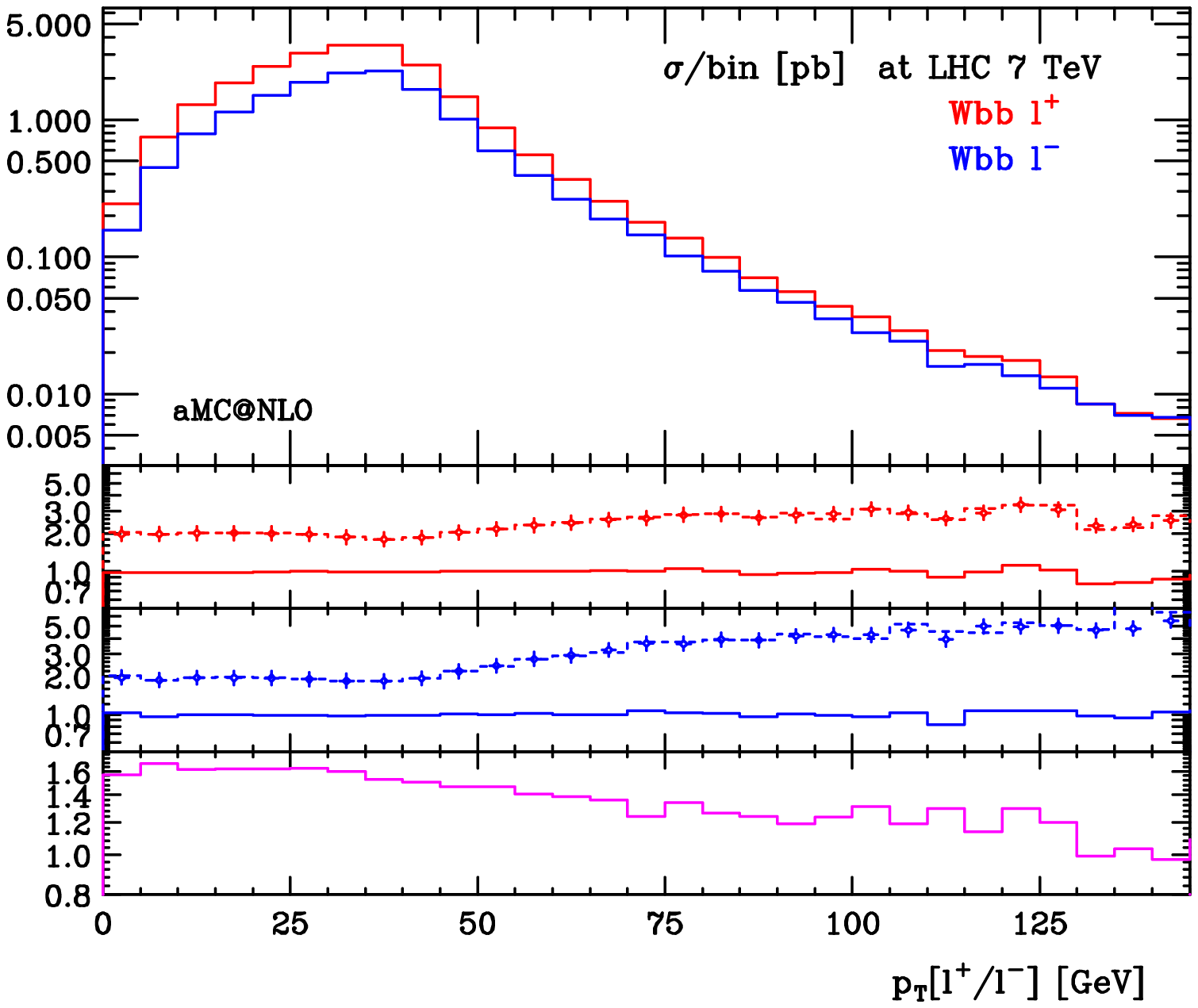}
\includegraphics[width=0.49\textwidth]{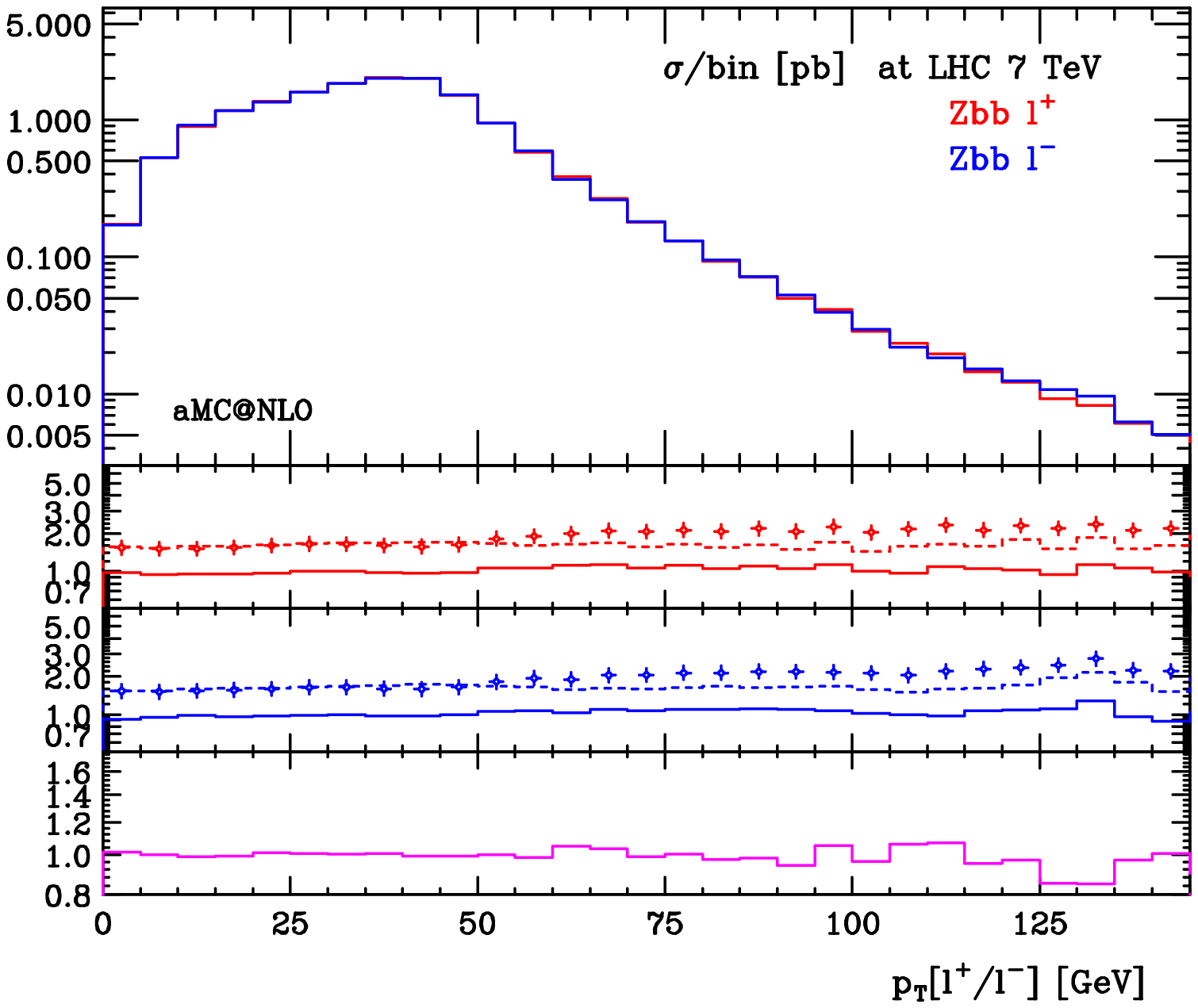}
\caption{Transverse momentum of the charged leptons in \Wbb\ (left panel) and
\Zbb\ (right panel) production, shown separately for positive and negative
charges.  The upper and middle insets follow the same patterns as those in
fig.~\protect\ref{fig:rates}. The lower inset (magenta solid histogram) is the
ratio of the \amcatnlo\ results relevant to positively-charged leptons 
over those relevant to negatively-charged ones.}
\label{fig:WZbb_pt_e}
\end{figure}

\begin{figure}[t]
\centering
\includegraphics[width=0.49\textwidth]{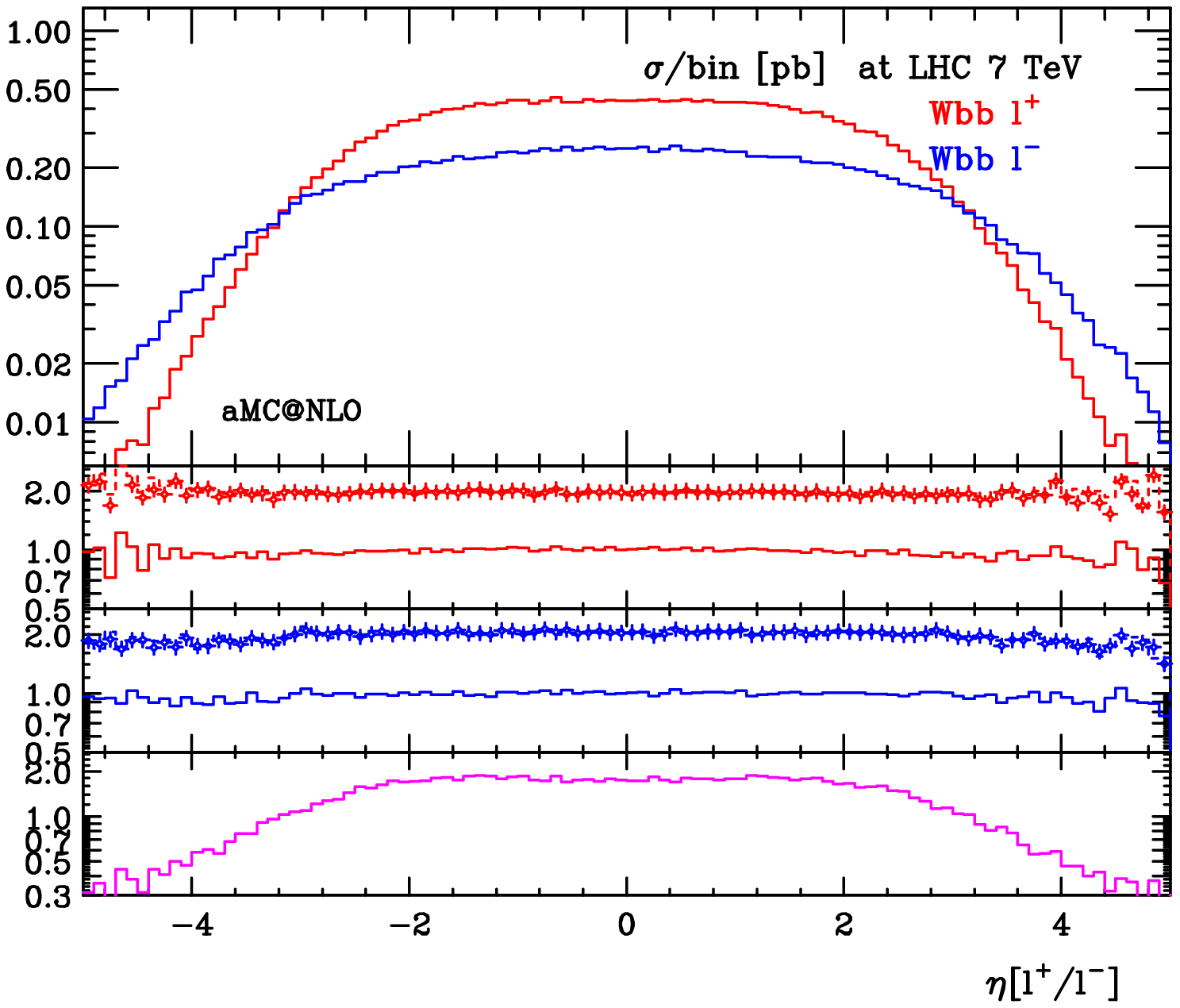}
\includegraphics[width=0.49\textwidth]{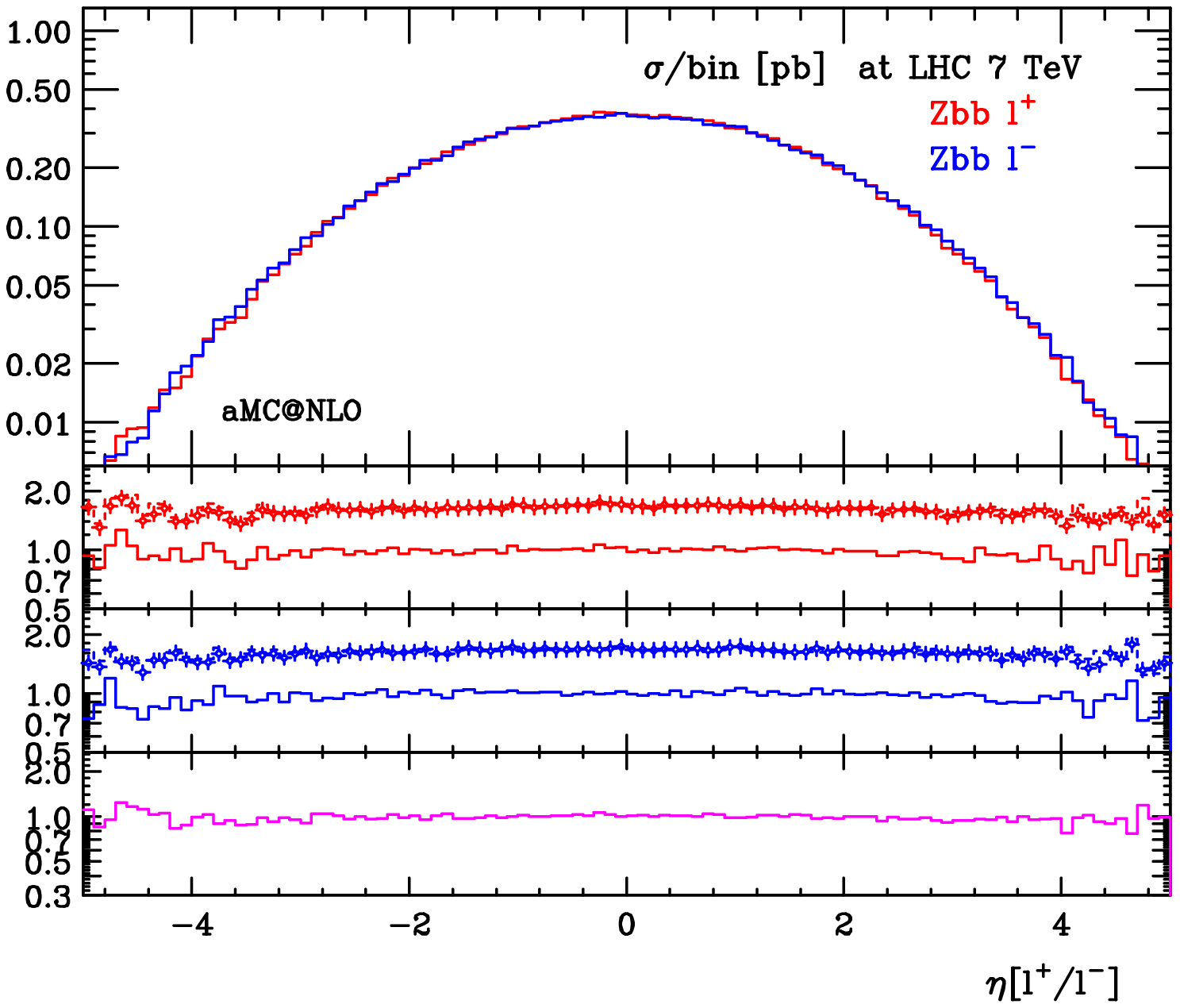}
\caption{As in fig.~\protect\ref{fig:WZbb_pt_e}, for the pseudorapidity of 
the charged leptons.}
\label{fig:WZbb_eta_e}
\end{figure}

In figs.~\ref{fig:WZbb_pt_e} and~\ref{fig:WZbb_eta_e} the transverse momenta
and the pseudorapidities of the charged leptons are shown separately according 
to their electric charges. In the two upper insets we have used the same 
patterns and conventions as in fig.~\ref{fig:rates} -- these will
be used throughout this paper.
In the case of \Wbbs\ production, the effects of the NLO corrections are 
especially pronounced at large $\pt$'s, where they are the signal
of new partonic subprocesses opening up at this order, and in particular
of those which include an initial-state gluon, such as $qg$. Results after
matching with showers consistently show a similar behaviour. 
The same large enhancement is not present in the case of $Z$ production,
which receives gluon-initiated contributions already at the LO; again,
this trend is seen also after matching with showers.
The lowest insets (solid magenta curves) show the ratios of the \amcatnlo\
results relevant to positively-charged leptons over those relevant to 
negatively-charged ones. In the case of \Wbbs\ production the behaviour is
similar to what has been described recently in ref.~\cite{Bern:2011ie}, while
for \Zbbs\ this distribution is flat, as expected.

\begin{figure}[t]
\centering
\includegraphics[width=0.50\textwidth]{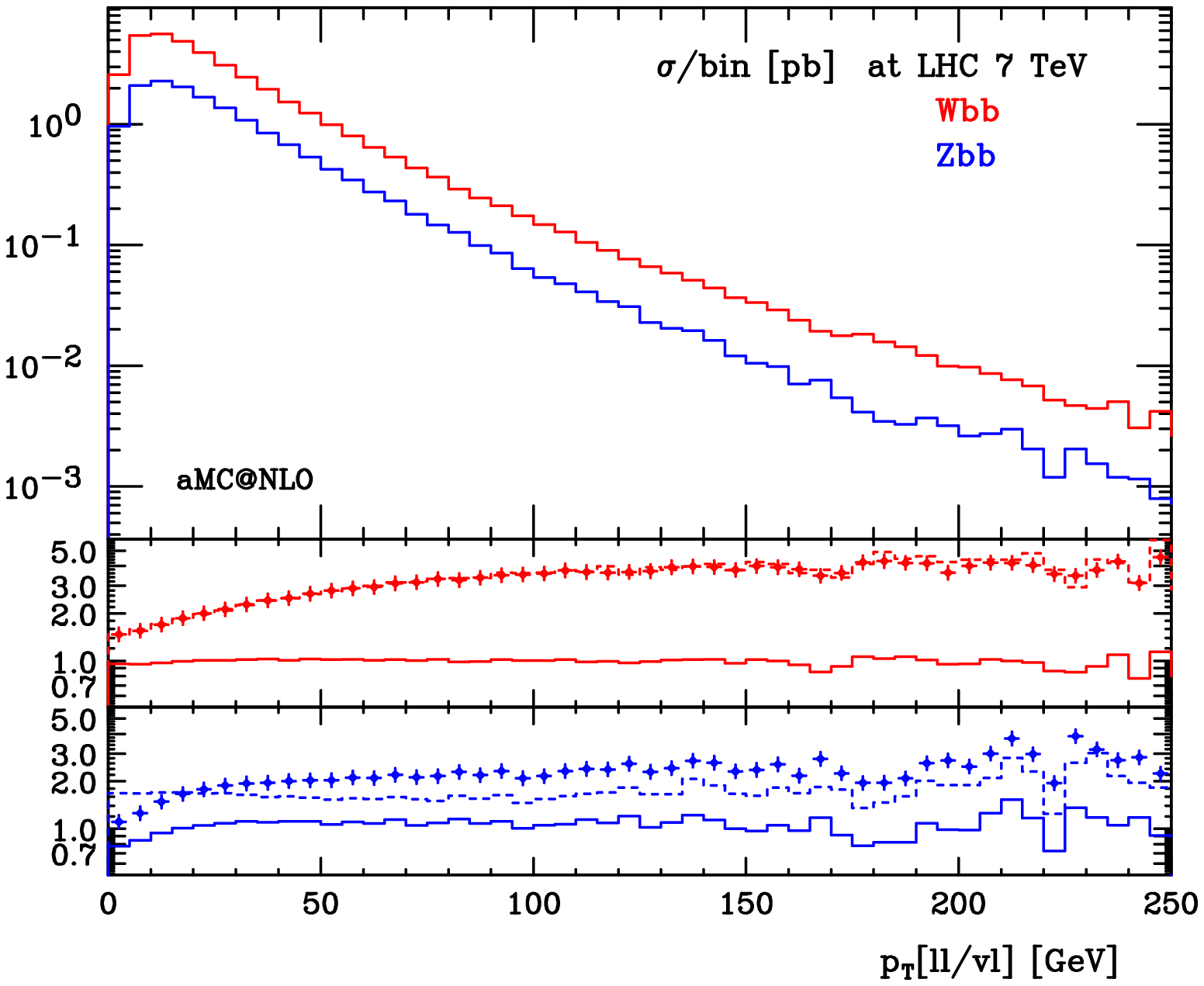}
\includegraphics[width=0.49\textwidth]{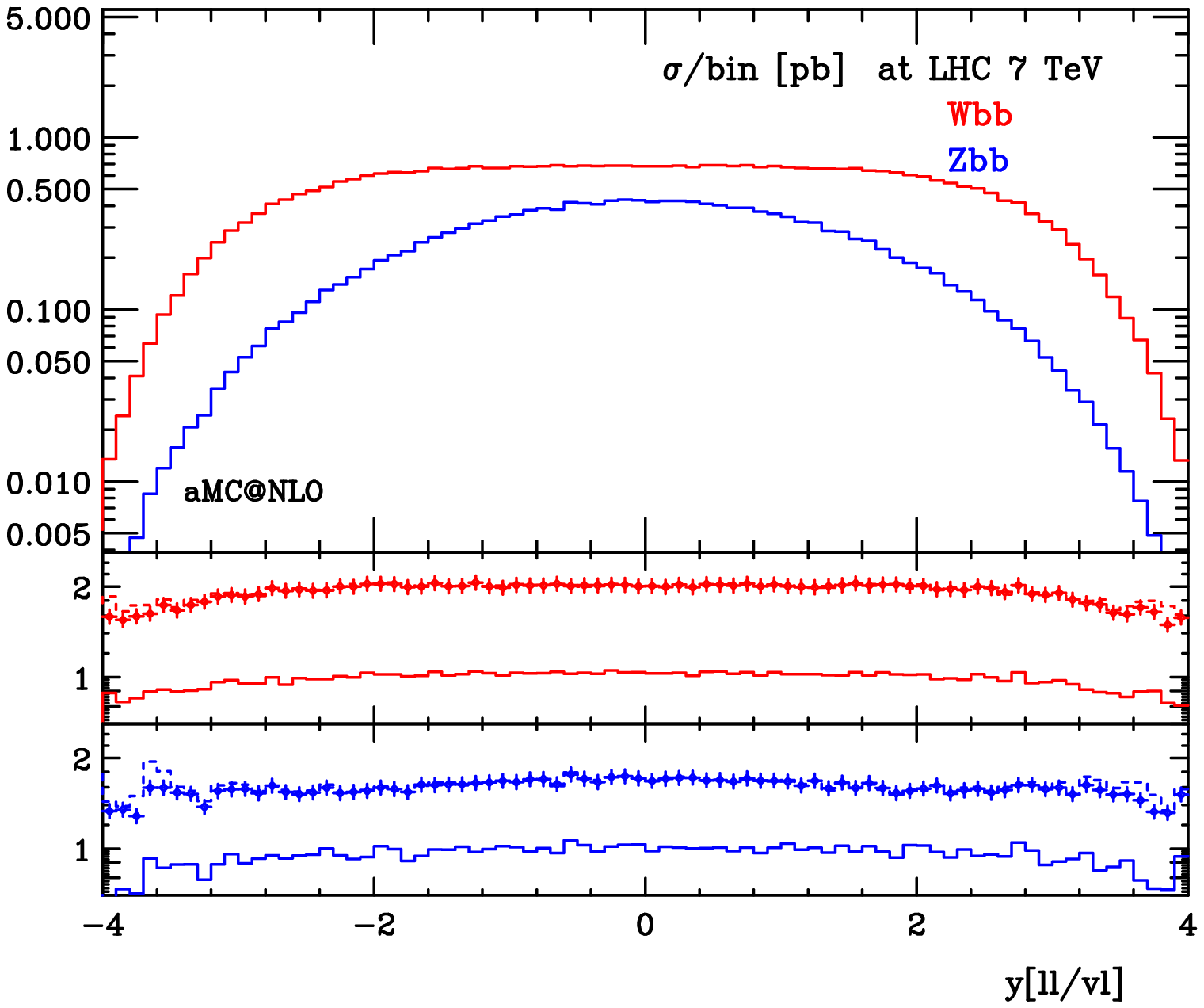}
\caption{Transverse momentum (left panel) and rapidity (right panel)
of the $\ell\nu$ and $\ell^+\ell^-$ pairs (i.e.~of the virtual $W$ and 
$Z$ bosons respectively) in \Wbb\ and \Zbb\ production.
The insets follow the same patterns as those in fig.~\protect\ref{fig:rates}.}
\label{fig:ptwz}
\end{figure}

A complementary aspect of the different parton luminosities that 
contribute to \Wbbs\ and \Zbbs\ production can be appreciated
by looking for example at the transverse momentum distributions of the 
$\ell\nu$ and $\ell^+\ell^-$ pairs (i.e.~of the virtual $W$ and 
$Z$ bosons respectively), shown in the left panel of fig.~\ref{fig:ptwz}.
In the case of \Wbb\ final states, the \amcatlo\ and LO results 
are very close to each other, which is not the case for \Zbb\
production. This is due to the fact that the $gg$-initiated channel
in the latter case is responsible for much more QCD radiation in MC-based
simulations than the $q\bar{q}$ channel (the latter being identical to
the mechanism that induces \Wbbs\ production). This larger amount of 
radiation hardens the virtual-$Z$ $\pt$ spectrum predicted by \amcatlo, 
making it more similar to the \amcatnlo\ result than in the case of 
\Wbbs\ production. The NLO $\pt$ spectra of the virtual vector bosons are
closer to the \amcatnlo\ results, because at that order one does get
contributions from gluon-initiated channels to both \Wbb\ and \Zbb\ final
states, and thus the relative changes obtained when matching with parton
showers have a milder impact. This is an example of the ``stabilizing''
pattern that one observes when higher-order perturbative results are taken
into account. On the other hand, the rapidity distributions of the lepton
pairs do not change significantly under QCD radiation, as is shown in the 
right panel of fig.~\ref{fig:ptwz}.

We remark that we do not find any significant enhancement in the 
large-$\pt$ tails of vector bosons when going from the NLO to the 
\amcatnlo\ predictions, at variance with the POWHEG \Wbbs\ result of 
ref.~\cite{Oleari:2011ey} which necessitates an ad-hoc {\em perturbative} 
tuning for this reason. It should further be stressed that the
almost perfect coincidence between the \amcatlo\ and LO results 
for $\pt(\ell\nu)$ may not occur by simply changing the tuning
of the shower parameters in \herwig: indeed, we have verified that
the spectrum predicted by \pythia~6 is slightly different w.r.t.~the
\amcatlo\ or LO ones. It is also interesting to notice that the features
discussed here and that affect the low- and intermediate-$\pt$ regions 
of lepton pairs are not visible in the case of the individual lepton $\pt$
spectra, fig.~\ref{fig:WZbb_pt_e}.  We have verified that kinematical
correlations are such that, for a fixed small or intermediate value of
$\pt(\ell^\pm)$, one integrates over a $\pt(\ell^+\ell^-)$ range that causes
the local differences between the \amcatlo\ and LO results for the latter
transverse momentum to be averaged out.

\begin{figure}[t]
\centering
\includegraphics[width=0.49\textwidth]{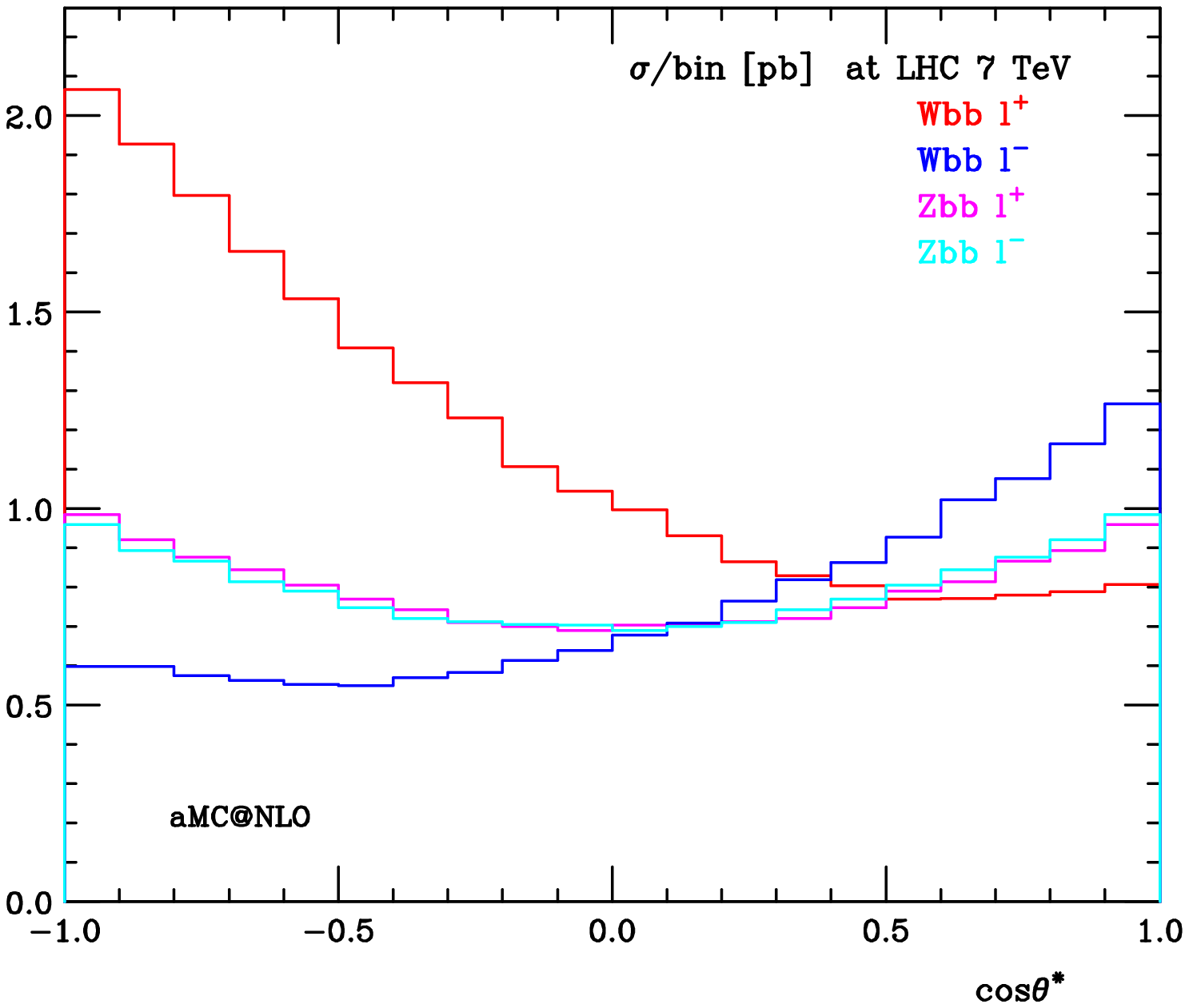}
\includegraphics[width=0.49\textwidth]{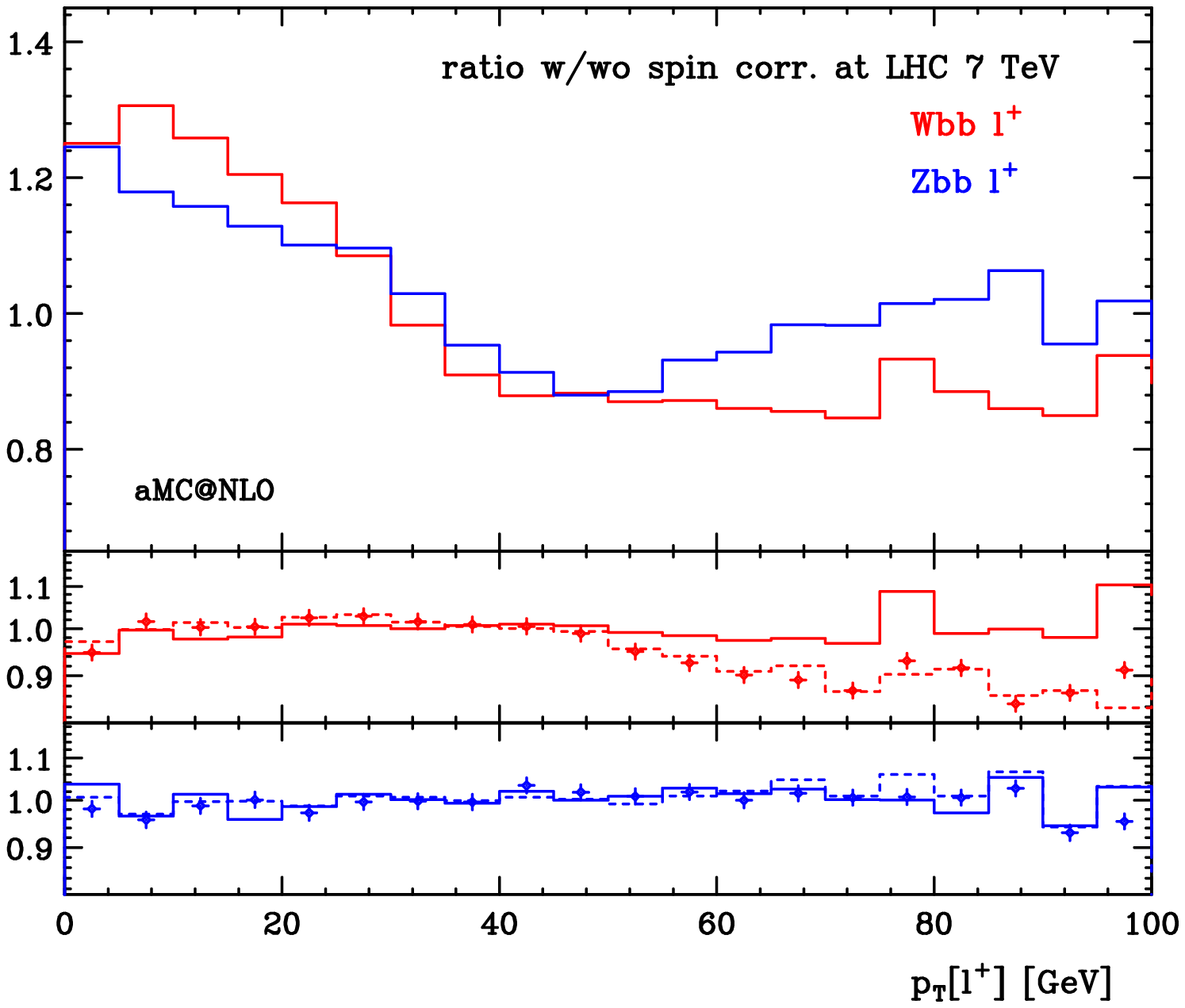}
\caption{Left panel: $\cos\theta^*$ distribution of final-state
charged leptons for different charges. All histograms have been 
obtained with \amcatnlo. See the text for the observable definition. 
Right panel: ratio of the results for the $\pt$ of the positively-charged 
lepton over the same quantity computed by neglecting production spin 
correlations. 
The insets follow the same patterns as those in fig.~\protect\ref{fig:rates}.}
\label{fig:costhstar}
\end{figure}

The left panel of fig.~\ref{fig:costhstar} presents the $\cos\theta^*$
distribution, computed by separating the positively- and negatively-charged
lepton contributions.
We remind the reader that such an observable is defined as the cosine of the
angle between the chosen charged lepton, and the direction of flight of the
parent vector boson, in the rest frame of the latter.  Clearly visible are the
strong angular correlations and charge asymmetry in the \Wbb\ case. For
\Zbb\ production such correlations and asymmetries, while present, 
are much milder, and likely not observable in a real experiment. 
For both processes, the \amcatnlo\ results are basically identical
to those of \amcatlo, NLO, and LO, and thus we refrain from showing
the latter here.

In the right panel of fig.~\ref{fig:costhstar}, where we consider only 
leptons with positive electric charge to be
definite, we plot the ratio of the lepton transverse momentum over the same
quantity, obtained by imposing a phase-space (i.e., flat) decay of the parent
vector boson; hence, this ratio is a measure of the impact of
spin correlations on the inclusive-lepton $\pt$. 
We see that differences between correlated and uncorrelated decays
can be as large as 20\%, and vary across
the kinematical range considered. This confirms that the inclusion of
spin-correlation effects is necessary when an accurate description of the
production process is required. We stress again that our computations feature
spin correlations exactly at the matrix-element level, including one-loop
ones. It is interesting to observe that, while in the case of \Zbbs\
production all four calculations give similar results (see the lower inset),
this happens in \Wbbs\ production only for $\pt(\ell^+)\lesssim 50$~GeV (see
the upper inset).  At $\pt$ values larger than this, \amcatnlo\ and NLO
predict ratios that differ from the corresponding \amcatlo\ and LO ones. Once
again, this is a manifestation of the significant impact of gluon-initiated,
NLO partonic processes on \Wbbs\ cross sections, and is consistent with 
the findings of ref.~\cite{Bern:2011ie} (relevant to the associated
production of $W$ bosons with light jets).
 
\begin{figure}[t]
\centering
\includegraphics[width=0.49\textwidth]{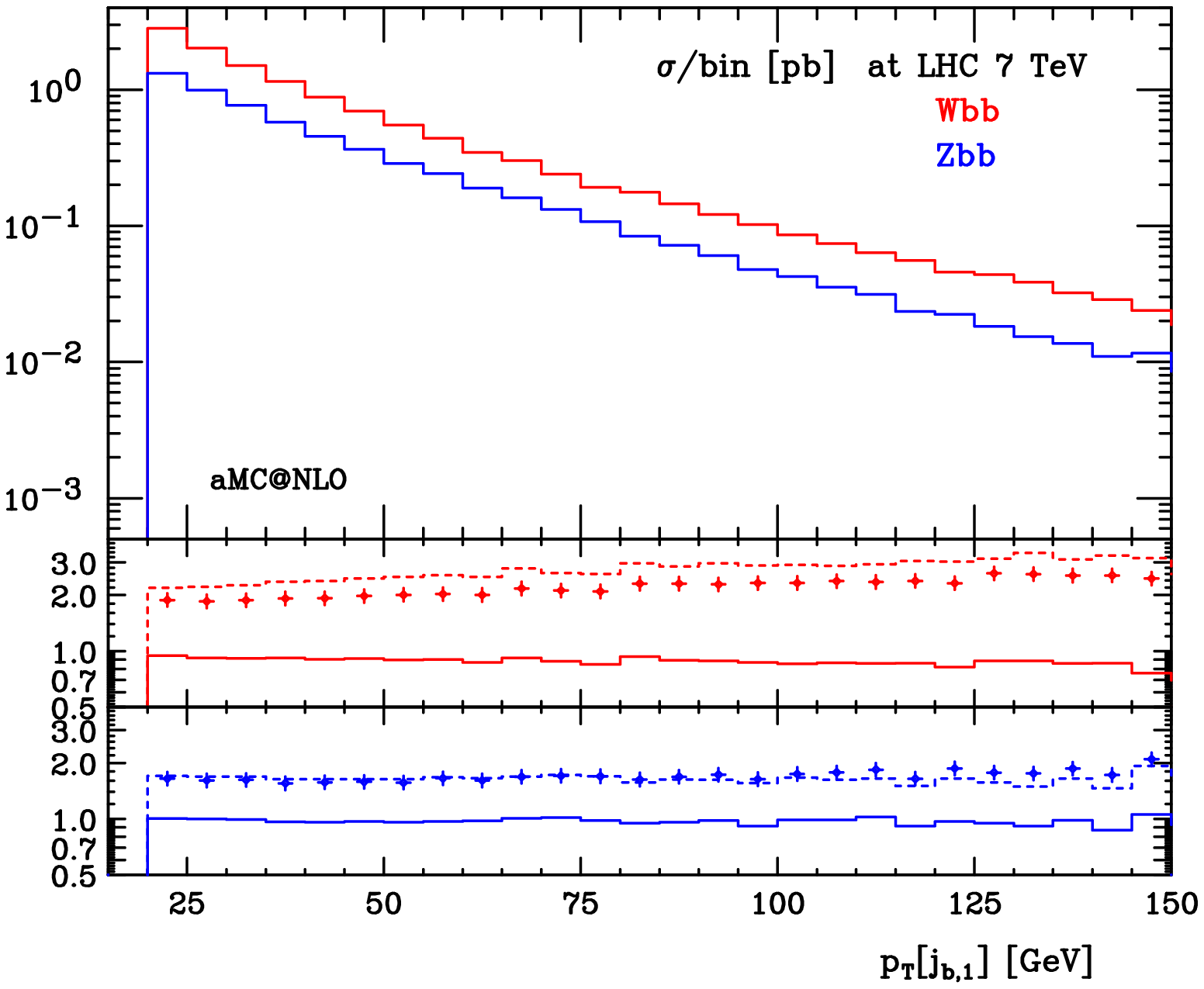}
\includegraphics[width=0.49\textwidth]{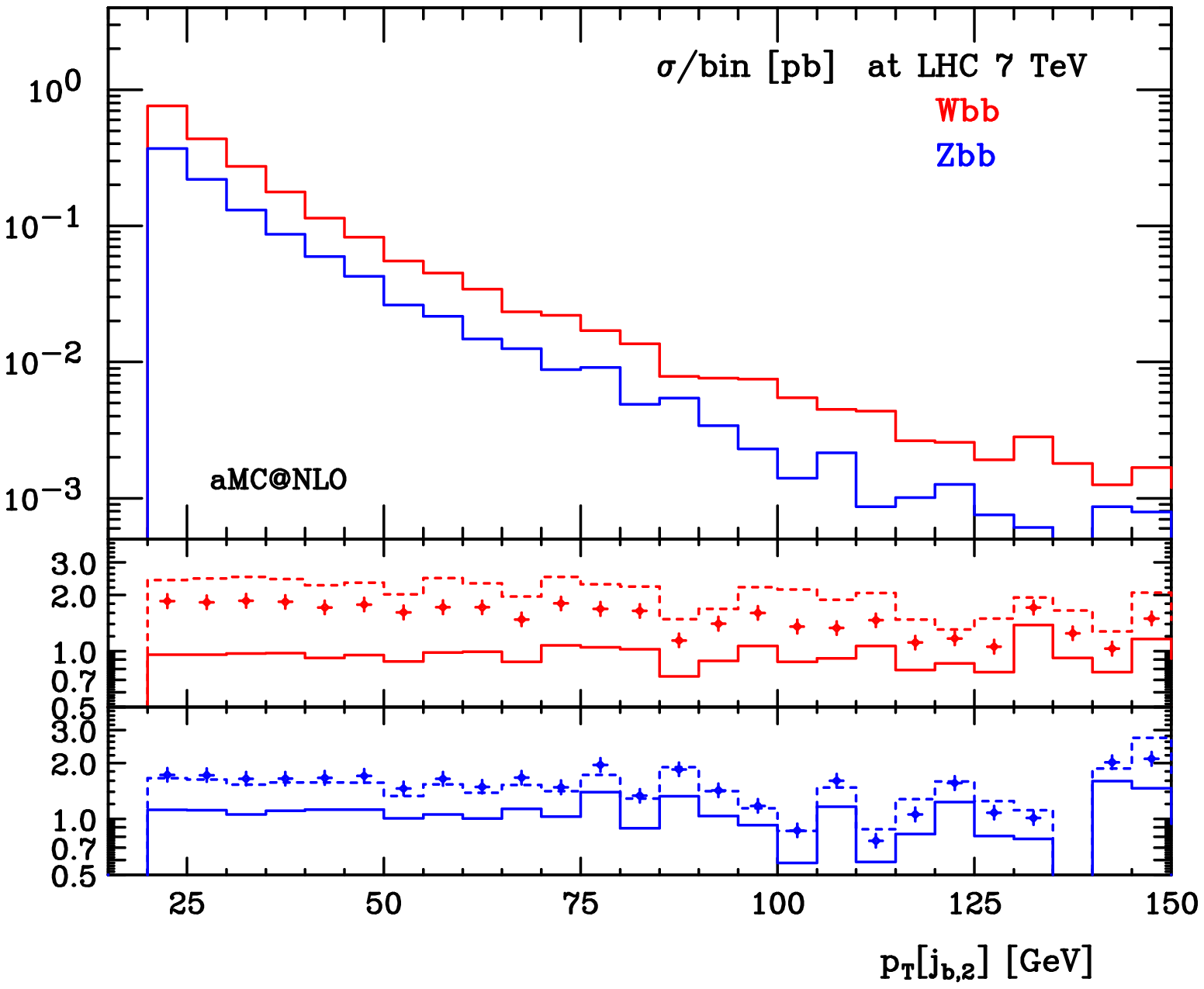}
\caption{Transverse momentum of the hardest (left panel) and
  second-hardest $b$-jet (right panel) in \Wbbs\ and \Zbbs\ production.
  The insets follow the same patterns as those in
  fig.~\protect\ref{fig:rates}.}
\label{fig:pt_bj1}
\end{figure}

\begin{figure}[t]
\centering
\includegraphics[width=0.49\textwidth]{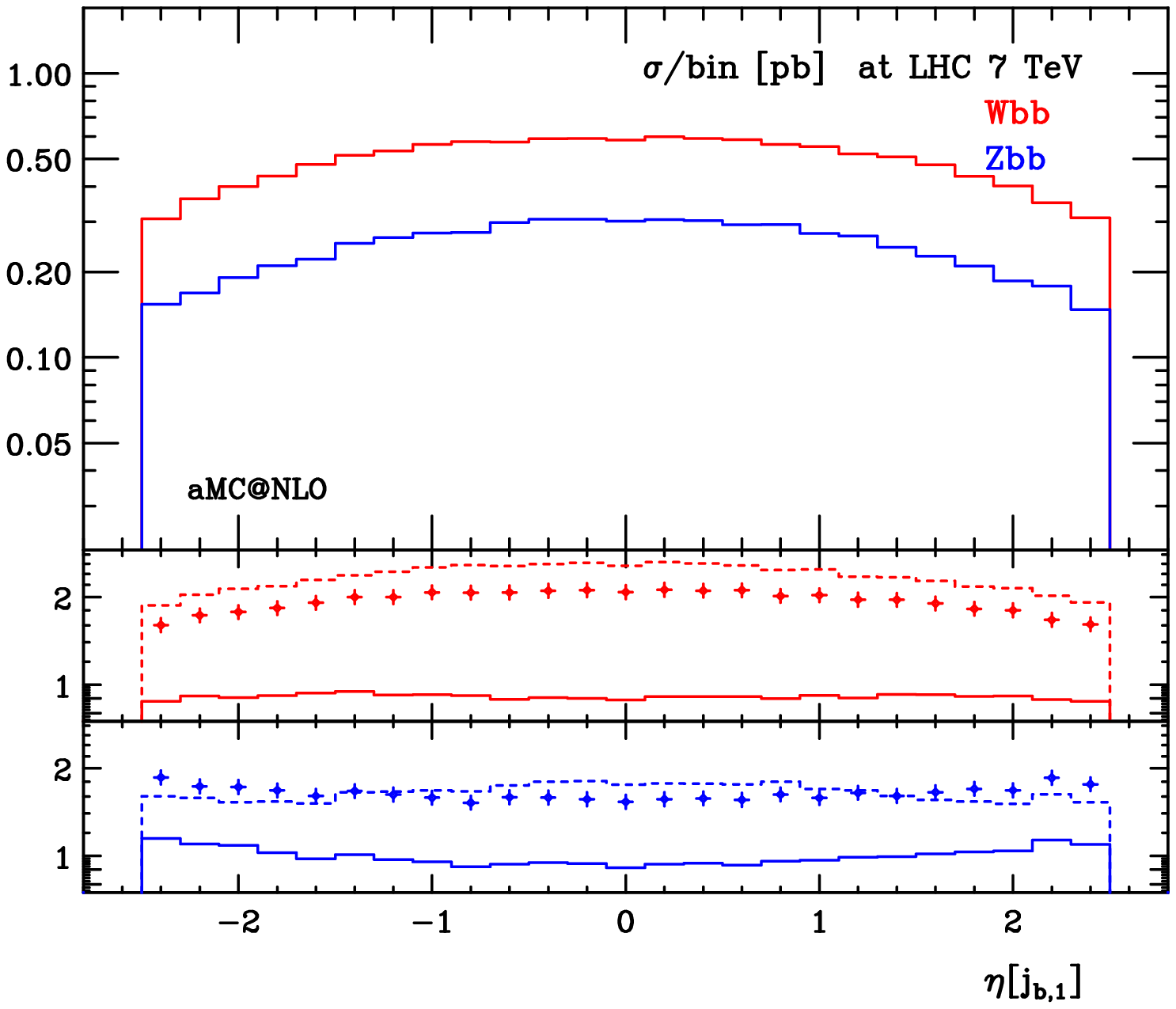}
\includegraphics[width=0.49\textwidth]{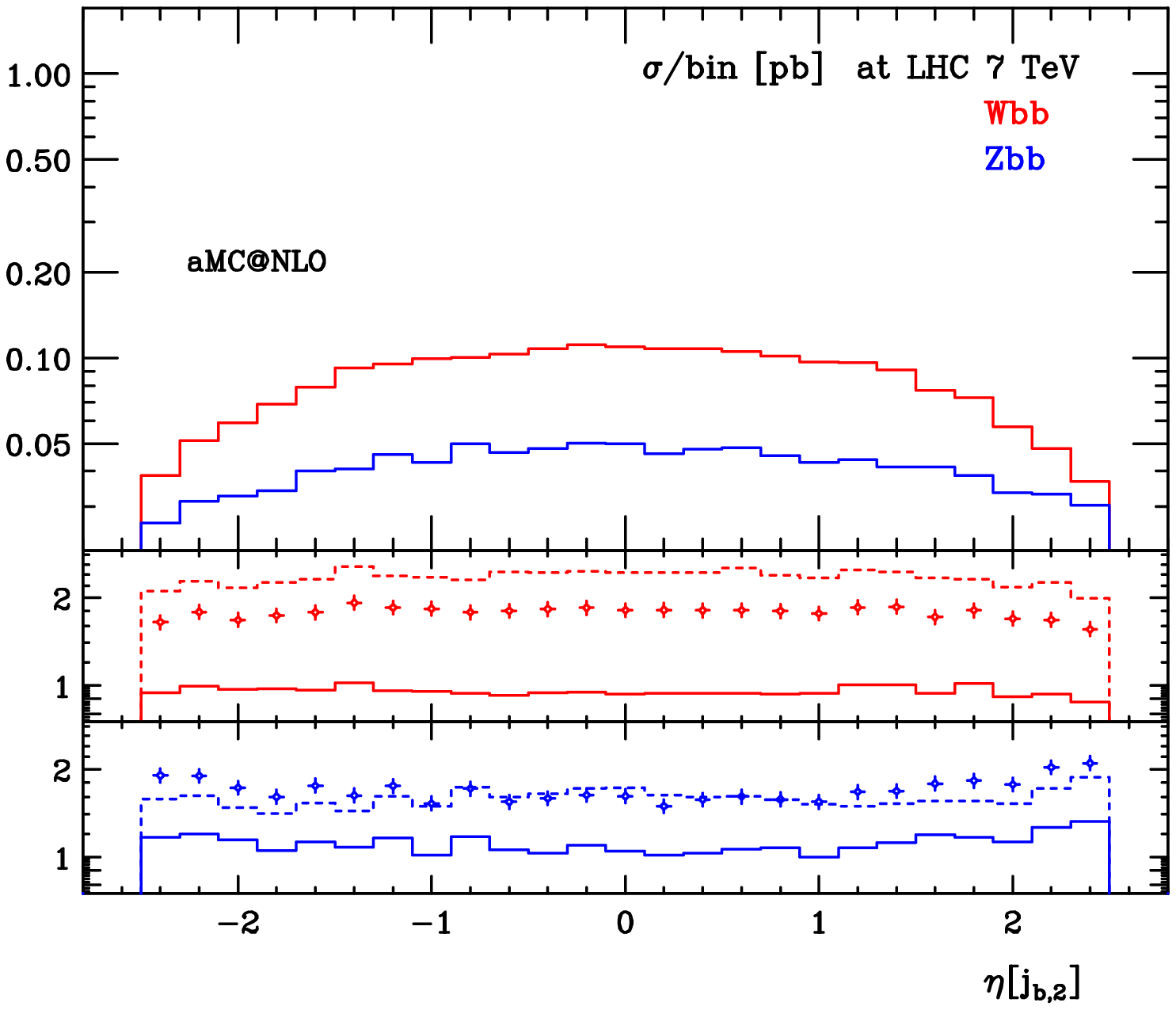}
\caption{As in fig.~\protect\ref{fig:pt_bj1}, for the pseudorapidity 
of the hardest and the second-hardest $b$-jet.}
\label{fig:eta_bj1}
\end{figure}

In figs.~\ref{fig:pt_bj1} and \ref{fig:eta_bj1} the transverse momenta and the
pseudorapidities of the two hardest $b$-jets are shown. Differences in 
normalisation are consistent with what we expect on the basis of inclusive 
$K$ factors; differences in shapes are typically small, but visible. We point 
out that for an event to contribute to the hardest-$b$-jet observables shown 
here it is sufficient that one $b$-jet be present in the event; the other 
$b$ quark emerging from the hard process can have arbitrarily small momentum.

\begin{figure}[t]
\centering
\includegraphics[width=0.49\textwidth]{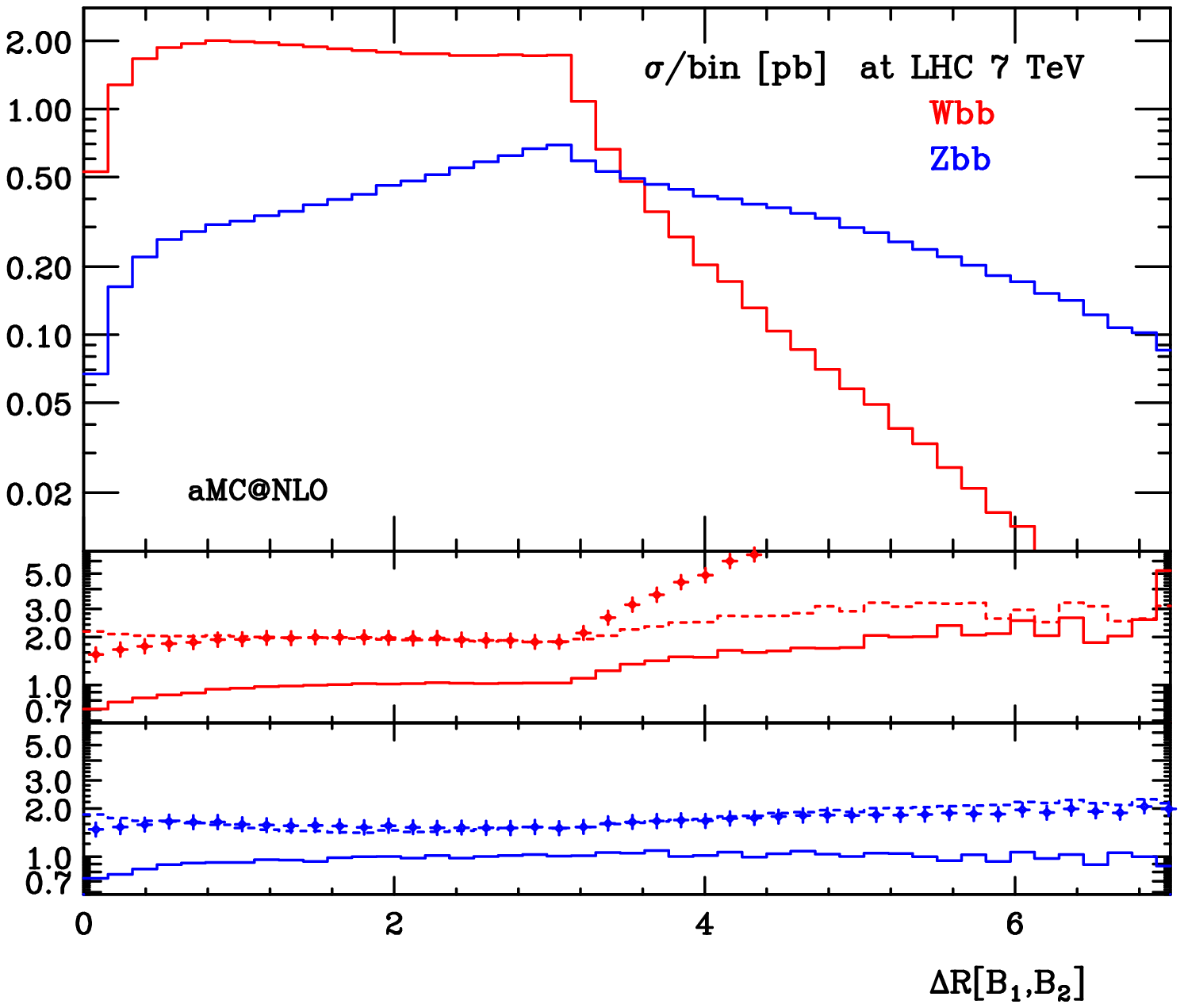}
\includegraphics[width=0.50\textwidth]{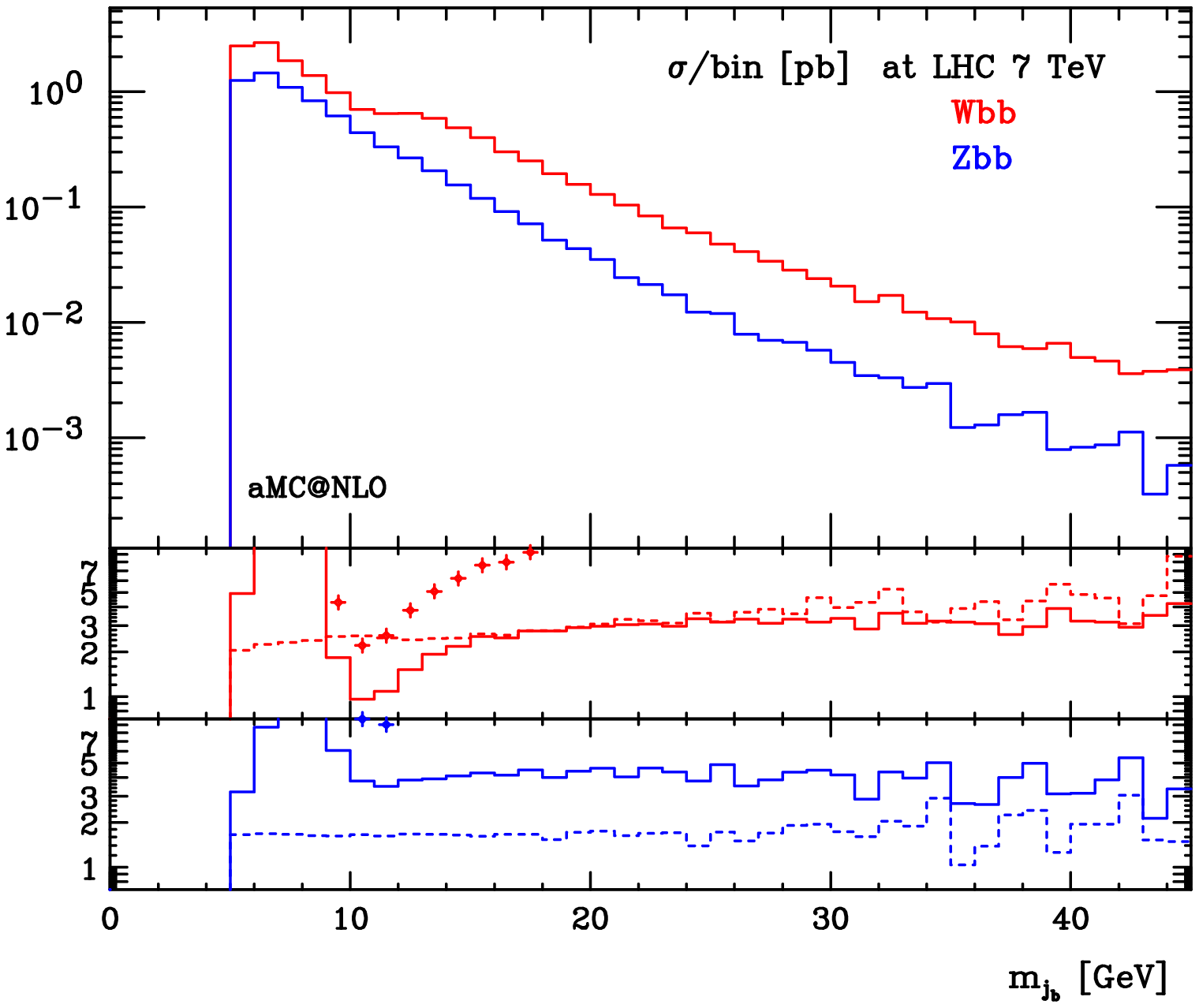}
\caption{Left panel: $\Delta R$ separation between the two hardest 
$b$-hadrons (\amcatnlo\ and \amcatlo) or the $b$ and $\bar{b}$ quarks
(NLO and LO) in the event. Right panel: invariant mass of the $b$-jets, 
inclusive over all $b$-jets in the event. 
The insets follow the same patterns as those in fig.~\protect\ref{fig:rates}.}
\label{fig:DelR_b12}
\end{figure}

In the left panel of fig.~\ref{fig:DelR_b12}, the $\Delta R$ separation 
between the two hardest $b$-hadrons (for the MC-based simulations)
or between the $b$ and $\bar{b}$ quarks (for the NLO and LO computations)
is shown. Differences between 
the \Wbbs\ and \Zbbs\ processes are manifest. In the
former case the two $b$'s originate from a final-state gluon
splitting, and they will thus tend to be quite close in pseudorapidity.
On the other hand, the two $b$'s in \Zbbs\ production can arise from 
the uncorrelated branchings of the initial-state gluons in the $gg$ channel,
and in this way they will naturally acquire a large separation in
pseudorapidity, which is directly related with large-$\Delta R$ values.
However, a $b\bar{b}$ pair arising from a final-state 
gluon branching can be easily separated in pseudorapidity by QCD radiation.
This is the reason why the parton-level LO result in the case of \Wbbs\ 
production is so different from the other three predictions (as shown
by the symbols in the upper inset). Both parton-level NLO (through
radiation present at the matrix-element level) and \amcatlo\ (through
radiation due to parton showers) results are in fact much closer
to \amcatnlo\ than parton-level LO result is. This does not happen
in the case of \Zbbs\ production, since as discussed before the $b$ and
$\bar{b}$ quarks can be well-separated in pseudorapidity already at the LO.
It should be stressed that the $b$-hadrons that contribute to
the $\Delta R$ separation shown in fig.~\ref{fig:DelR_b12} are not
subject to any lower cuts in $\pt$. Thus, one expects that the
effects of extra radiation be diminished when imposing a $\pt$ cut
or, which is equivalent, by studying the same distribution in the
case of $b$-jets. We have verified that this is indeed the case,
i.e.~that when a minimum-$\pt$ cut is imposed on the two $b$-hadrons
the pattern of NLO QCD corrections in \Wbbs\ production is
more similar to that observed in \Zbbs\ production.
This is another example of the possibility of testing detailed
properties of QCD radiation by considering low-$\pt$ events.
It should be clear that from the theoretical viewpoint such studies
can be sensibly performed only by retaining the full $b$-mass dependence.
 
The right panel of fig.~\ref{fig:DelR_b12} shows the mass of
the $b$-jets in the events. The observable is inclusive over all
$b$-jets, which implies that a given event may enter more than once in the
plot. Striking is the onset of the $bb$-jet contribution in the \Wbbs\
result around $m_{j_b}\approx 12$~GeV. In the case of \Zbbs\ production
this effect is almost invisible, a consequence of the fact that the fraction 
of events where a $b$-jet is actually a $bb$-jet is much smaller
than for \Wbbs\ production, see fig.~\ref{fig:rates}. The distribution 
discussed here measures the activity inside a jet, and one cannot expect
fixed-order parton-level results, where a jet consists of one or two
particles, to be particularly sensible in this case. In fact, we see that
fixed-order results are very different from MC-based ones. On the other
hand, the differences in shape when going from \amcatlo\ to \amcatnlo\ 
are small, in particular for \Zbbs\ production, as expected for 
observables which are insensitive to emissions at large relative $\pt$'s.
We also point out that the knee at $m_{j_b}\approx 12$~GeV would
appear as a feature of {\em gluon} jets if the $b$-jet definition of
ref.~\cite{Banfi:2006hf} were used. This stresses again the fact that,
at small and moderate $\pt$'s, the usual definition gives more intuitive
results. On the other hand, at large jet $\pt$'s the onset of the
$bb$-jet contribution to $m_{j_b}$ is largely smeared out.

\begin{figure}[t]
\centering
\includegraphics[width=0.49\textwidth]{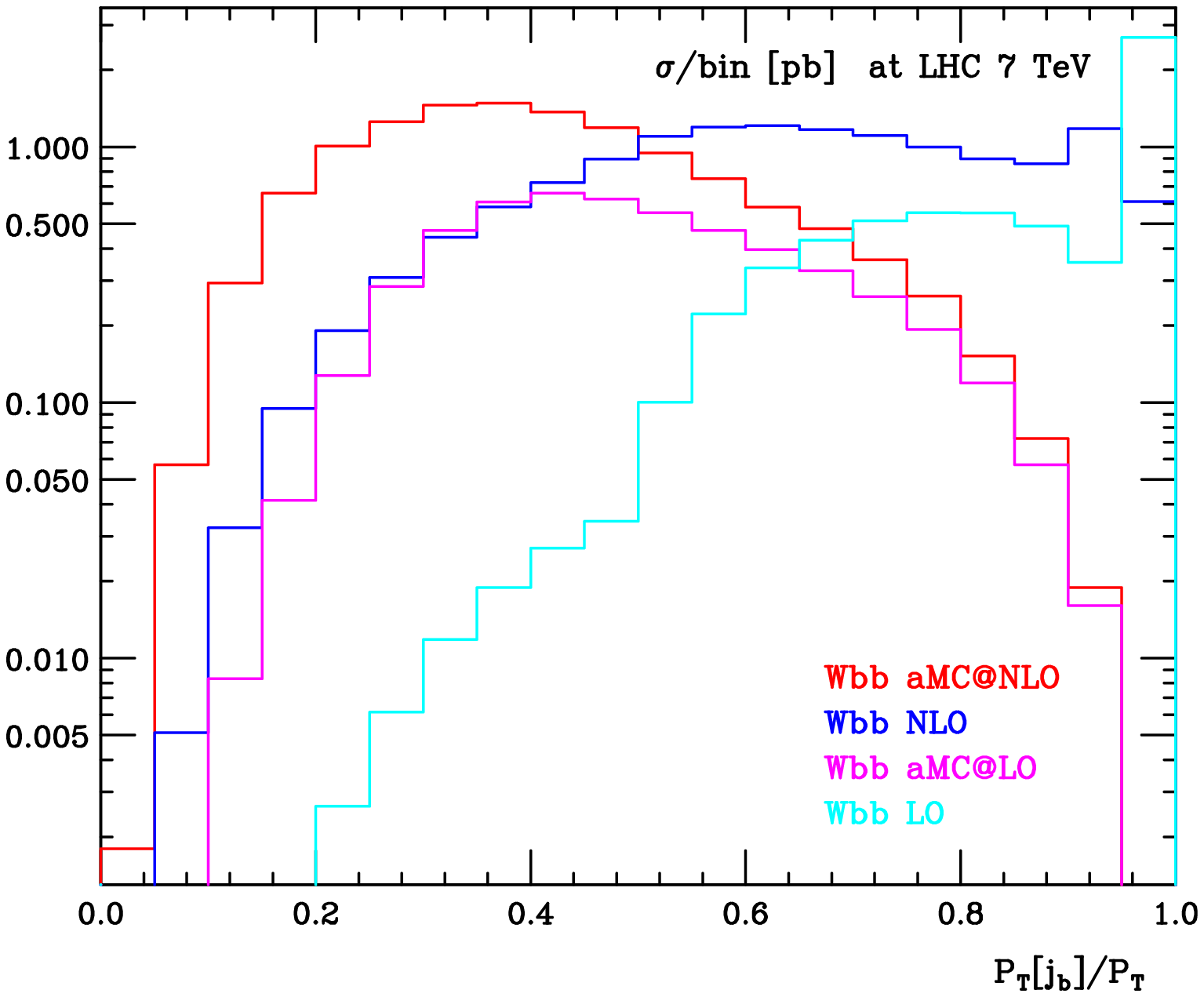}
\includegraphics[width=0.49\textwidth]{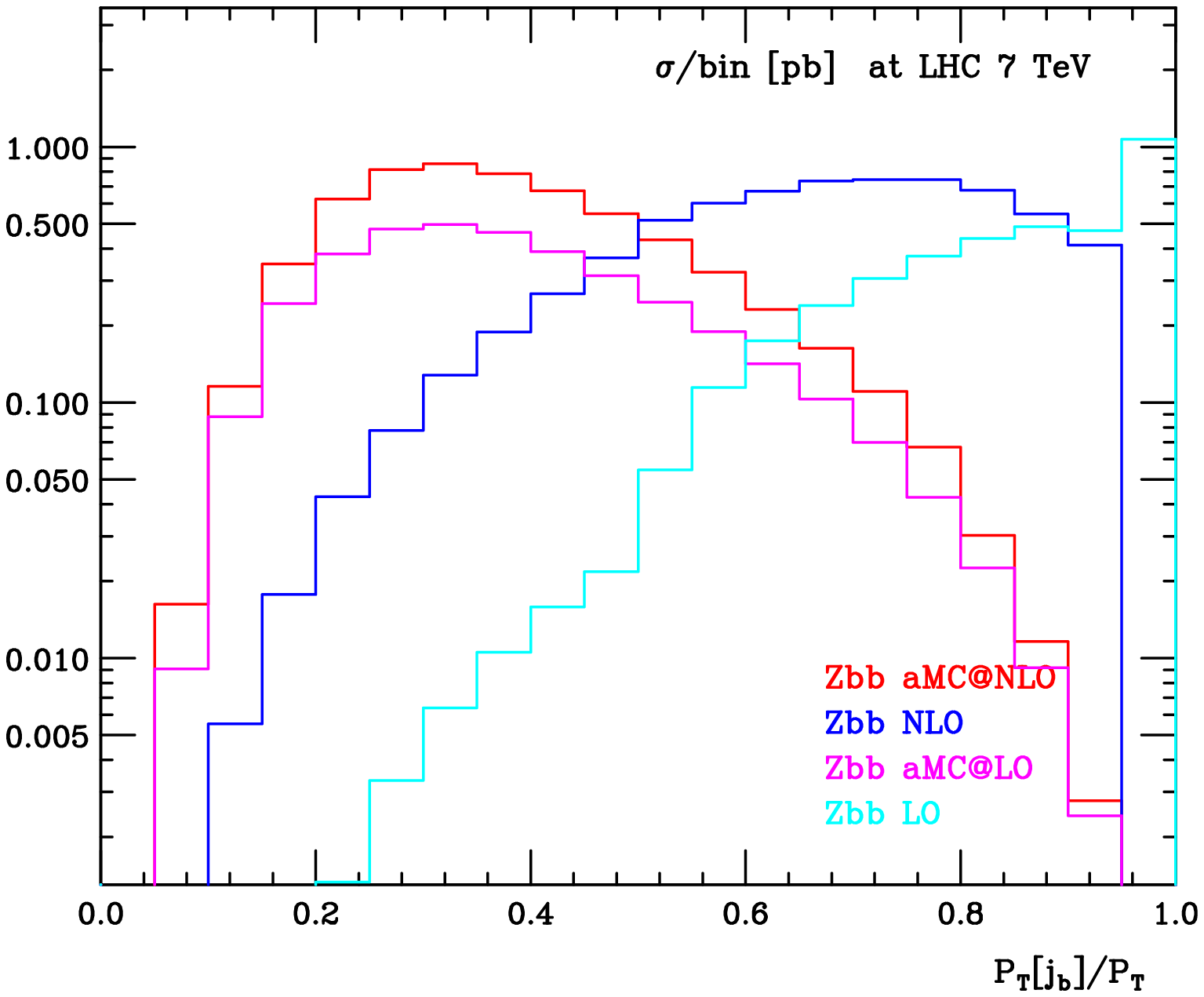}
\caption{Transverse momentum fraction carried by $b$-jets.
See the text for details.}
\label{fig:Etjb}
\end{figure}
In fig.~\ref{fig:Etjb} we show the ratio of the total transverse 
momentum $\Pt[j_b]$ of $b$-jets, over the total transverse hadronic
momentum $\Pt$\footnote{We stress that $\Pt$ is defined without including
the underlying event and pile-up contributions.}. In the context of 
parton-level computations, by ``hadrons'' we simply understand QCD partons.
At the parton-level LO, the configurations with one $bb$-jet or
with two $b$-jets (each of which contains one $b$ quark)
give contribution at $\Pt[j_b]/\Pt=1$. Configurations
with one $b$-jet that contains only one $b$ quark contribute to
$0.5< \Pt[j_b]/\Pt <1$ if the other $b$ quark has 
$\pt<20$~GeV (i.e., it is softer than a jet is required to be), while
values $\Pt[j_b]/\Pt<0.5$ can be obtained when the other $b$ quark 
has $\pt>20$~GeV and ${\abs{\eta(b)}}>2.5$ (i.e., it is outside
the $b$-jet tagging region in pseudorapidity). 
What was said above implies that $\Pt[j_b]/\Pt=1$
is an infrared-sensitive region, which gives rise to Sudakov logarithms
at higher order; this explains the behaviour of the parton-level NLO results 
there. Furthermore, the LO contributions to the \mbox{$\Pt[j_b]/\Pt<0.5$}
region decrease when increasing the maximum-pseudorapidity cut on jets. This 
is only marginally the case at the NLO (because of the presence of a hard 
light parton in the real-emission contributions), which explains the 
longer tail of the latter results w.r.t.~the LO ones.
The arguments above obviously do not apply to the context of an
event generator; this is confirmed by the similarity of the
\amcatnlo\ and \amcatlo\ results.
Firstly, at $\Pt[j_b]/\Pt=1$ Sudakov logarithms are properly resummed.
Secondly, the extra radiation generated by parton showers implies that
quite a few hadrons will lie outside $b$-jets, hence shifting further
the $\Pt[j_b]/\Pt$ results to the left of those relevant to parton-level
NLO computations. This shift is also present when passing from the
\amcatlo\ to the \amcatnlo\ predictions in \Wbbs\ production, while
in the case of \Zbbs\ production these two results are very similar
(up to an overall rescaling by the inclusive $K$ factor). We are finding
here the same pattern already discussed for a few observables in this
paper. Namely, the opening of gluon-initiated partonic channels at the
NLO in \Wbbs\ production implies a richer hadronic activity w.r.t.~the
corresponding LO case, which is only marginal in the case of $Zbb$
production owing to the dominance of the $gg$ channel already at the LO there.
Hence, the relative enhancement of the hadronic activity outside the $b$-jets
when going from \amcatlo\ to \amcatnlo\ is stronger for \Wbbs\ production
than is for \Zbbs\ production.

\begin{figure}[t]
\centering
\includegraphics[width=0.6\textwidth]{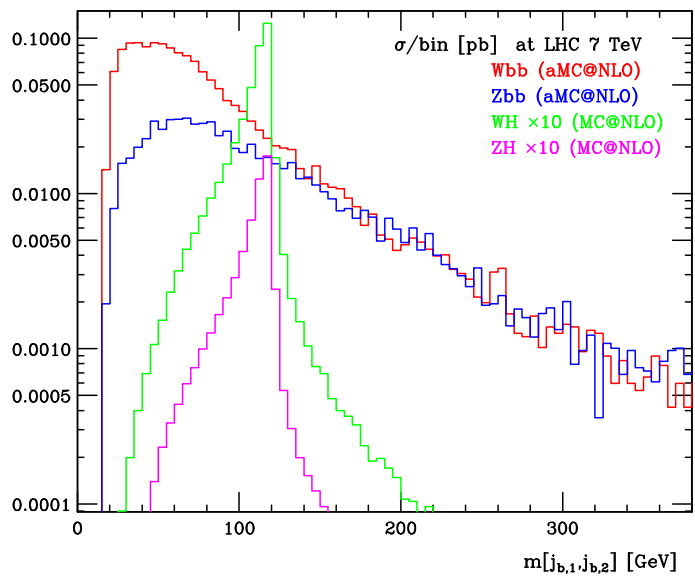}
\caption{Invariant mass of the pair of the two leading $b$-jets.
$WH(\to \ell\nu b\bar{b})$, $ZH(\to\ell^+\ell^- b\bar{b})$, \Wbb,
and \Zbb\ results are shown, with the former two rescaled by a
factor of ten.}
\label{fig:M_jb12_h}
\end{figure}
Finally, as a simple application to Higgs searches of the calculations
presented in this paper,
we show in fig.~\ref{fig:M_jb12_h} the invariant mass of the two leading 
$b$-jets in $WH(\to \ell\nu b\bar{b})$, $ZH(\to\ell^+\ell^- b\bar{b})$, 
\Wbb, and \Zbb\ events. The former two processes (the ``signal'') have 
been simulated with MC@NLO~\cite{Frixione:2002ik}\footnote{In the process 
of validating \amcatnlo, we had checked that it gave results identical 
to MC@NLO for all the processes implemented in the latter. Hence, we 
could have equally well employed \amcatnlo\ to simulate the signal here.}, 
with a Higgs mass $m_H=120$~GeV.
The tail at \mbox{$m[j_{b,1},j_{b,2}]>m_H$} is due to the fact that
the jet momenta are typically larger than those of the $b$-hadrons they
contain, owing to the contributions of other final-state hadrons emerging 
from initial-state showers. This is compensated by the fact that the 
$b$-hadron momenta are only a fraction of those of their parent $b$ quarks, 
the complementary fraction being lost to radiation which may end up
outside the jets. These two effects
smear the Higgs peak. Furthermore, in some events the $b$ quarks entering 
the two hardest $b$-jets do not arise from the Higgs decay, but from 
a $g\to b\bar{b}$ branching in the shower phase. Although rare indeed,
these events may result in invariant masses much larger than the
Higgs pole mass. The comparison given here is just an example of an 
analysis in which both the signal and its irreducible backgrounds can 
be computed at the same precision with (a)\mcatnlo, improving upon both 
fixed-order and LO-based Monte Carlo descriptions.

\section{Conclusions and outlook}
\label{sec:con}
In this work we have presented results for the \Wbb\ and \Zbb\
production processes, accurate to the NLO in QCD and that include the matching
to parton showers according to the MC@NLO formalism.
Our approach is fully general, completely automated, and opens the way
to performing comparisons with experimental data from the Tevatron and the LHC
at the highest theoretical accuracy attainable nowadays.

By studying a limited but representative set of observables, we have shown 
that several are the elements to be kept into account in order to achieve
reliable and flexible predictions for this class of processes: spin
correlations of the final state leptons emerging from the decays of the 
vector bosons, heavy-quark mass effects, and a realistic description
of the final states, obtained thanks to the interface with a shower and
hadronisation program. As we have seen, NLO QCD corrections have a
highly non-trivial impact, since they lead not only to large enhancements 
of total rates, but also to significant changes in the shapes 
of distributions. In this respect, the opening at the NLO of new 
partonic channels, and in particular of those involving gluons, plays 
a fundamental role. In general and apart from well-understood cases in
which pure perturbative results are not meaningful, one observes that 
at the NLO level fixed-order and MC-based results are closer to each other
than the corresponding LO ones. This is in keeping with naive
expectations based on perturbation theory, and it is significant 
in that it shows that the very large corrections affecting the processes
considered here do not pose problems when the matching with parton
shower Monte Carlos is carried out according to the \mcatnlo\ method.

Thanks to the \amcatnlo\ implementation, several QCD issues interesting on
their own can now be addressed.  One example over all is the study of NLO
corrections, mass effects and radiation pattern in final-state gluon 
splitting, for which \Wbbs\ production offers a particularly clean 
environment. Gluon splitting in the initial state and the role of the 
$b$ PDF (and therefore of different schemes for the predictions of total and
differential cross sections) can be assessed by considering \Zbbs\ production.
The outcome of such a study can then be applied to the $Hb \bar b$ case.  In
particular, available predictions in the five-flavour scheme at
the NNLO for the fully-inclusive production of a $Z$ in association with
bottom quarks~\cite{Maltoni:2005wd}, and for $Z$+1 $b$-jet at the
NLO~\cite{Campbell:2003dd}, can now compared with our four-flavour-scheme
results. In addition, QCD radiation effects on high-$\pt$ $b\bar b$ pairs,
which can be merged into one jet, are also of interest in boosted-Higgs
searches~\cite{Butterworth:2008iy}. Finally, spin correlation effects may also
be investigated to gather more insight on the production mechanisms in QCD,
and possibly to distinguish them from other competing hard reactions, such as
double-parton scatterings. We plan to address some of the above issues in
detail in the near future.

We conclude by pointing out that event files relevant to the processes
studied in this paper (as well as to others) are publicly available at 
{\tt http://amcatnlo.cern.ch}. Work to make the use of \amcatnlo\ public 
from the same site is in progress.

\section{Acknowledgments}
S.F. would like to thank Gavin Salam for useful discussions.
This research has been supported by the Swiss National Science
Foundation (NSF) under contract 200020-126691, by the Belgian IAP Program,
BELSPO P6/11-P and the IISN convention 4.4511.10, by the Spanish Ministry 
of education under contract PR2010-0285, and in part by
the US National Science Foundation under grant No.~NSF PHY05-51164. 
F.M. and R.P. thank the financial support of the MEC 
project FPA2008-02984 (FALCON).

\bibliography{Vbb}


\end{document}